\documentclass[12pt,preprint]{aastex}

\def\hal{H$\alpha$}
\def\be{\begin{equation}}
\def\ee{\end{equation}}

\def\m{~$\mu$m}

\def\CI   {[\ion{C}{1}]}
\def\CII  {[\ion{C}{2}]}
\def\HI   {\ion{H}{1}}
\def\HII  {\ion{H}{2}}
\def\FeII {[\ion{Fe}{2}]}

\def\NeII {[\ion{Ne}{2}]}
\def\NeIII{[\ion{Ne}{3}]}
\def\NeV  {[\ion{Ne}{5}]}
\def\OI   {[\ion{O}{1}]}

\def\OIV  {[\ion{O}{4}]}

\def\SIII {[\ion{S}{3}]}
\def\SIV  {[\ion{S}{4}]}
\def\SiII {[\ion{Si}{2}]}

\def\Spitzer{{\it Spitzer}}
\def\2MASS{{\it 2MASS}}

\begin {document}

\title{The Spitzer Infrared Nearby Galaxies Survey: A High-Resolution Spectroscopy Anthology}

\author{D.A.~Dale\altaffilmark{1}, J.D.T.~Smith\altaffilmark{2}, E.A.~Schlawin\altaffilmark{1}, L.~Armus\altaffilmark{3}, B.A.~Buckalew\altaffilmark{4}, S.A.~Cohen\altaffilmark{1}, G.~Helou\altaffilmark{3}, T.H.~Jarrett\altaffilmark{3}, L.C.~Johnson\altaffilmark{1}, J.~Moustakas\altaffilmark{6}, E.J.~Murphy\altaffilmark{3}, H.~Roussel\altaffilmark{7}, K.~Sheth\altaffilmark{3}, S.~Staudaher\altaffilmark{1}, C. Bot\altaffilmark{8}, D.~Calzetti\altaffilmark{9}, C.W.~Engelbracht\altaffilmark{10}, K.D.~Gordon\altaffilmark{11}, D.J.~Hollenbach\altaffilmark{12}, R.C.~Kennicutt\altaffilmark{5}, S.~Malhotra\altaffilmark{13}}
\altaffiltext{1}{\scriptsize Department of Physics and Astronomy, University of Wyoming, Laramie, WY 82071; ddale@uwyo.edu}
\altaffiltext{2}{\scriptsize Department of Physics and Astronomy, University of Toledo, Toledo, OH 43606}
\altaffiltext{3}{\scriptsize Spitzer Science Center, California Institute of Technology, Pasadena, CA 91101}
\altaffiltext{4}{\scriptsize Embry-Riddle Aeronautical University, Prescott, AZ 86301}
\altaffiltext{5}{\scriptsize Institute of Astronomy, University of Cambridge, Cambridge, United Kingdom}
\altaffiltext{6}{\scriptsize Center for Cosmology and Particle Physics, New York University, New York, NY 10003}
\altaffiltext{7}{\scriptsize Institut d'Astrophysique de Paris, 75014 Paris, France}
\altaffiltext{8}{\scriptsize Observatoire Astronomique de Strasbourg, 67000 Strasbourg, France}
\altaffiltext{9}{\scriptsize Department of Astronomy, University of Massachusetts, Amherst, MA 01003}
\altaffiltext{10}{\scriptsize Steward Observatory, University of Arizona, Tucson, AZ 85721}
\altaffiltext{11}{\scriptsize Space Telescope Science Institute, Baltimore, MD 21218}
\altaffiltext{12}{\scriptsize NASA/Ames Research Center, Moffett Field, CA 94035}
\altaffiltext{13}{\scriptsize Department of Physics and Astronomy, Arizona State University, Tempe, AZ 85287}

\begin {abstract}
High resolution mid-infrared spectra are presented for 155 nuclear and extranuclear regions from the \Spitzer\ Infrared Nearby Galaxies Survey (SINGS).  The fluxes for nine atomic forbidden and three molecular hydrogen mid-infrared emission lines are also provided, along with upper limits in key lines for infrared-faint targets.  The SINGS sample shows a wide range in the ratio of \SIII~18.71\m/\SIII~33.48\m, but the average ratio of the ensemble indicates a typical interstellar electron density of 300--400~cm$^{-3}$ on $\sim$23\arcsec$\times$15\arcsec\ scales and 500--600~cm$^{-3}$ using $\sim$11\arcsec$\times$9\arcsec\ apertures, independent of whether the region probed is a star-forming nuclear, a star-forming extranuclear, or an AGN environment.  Evidence is provided that variations in gas-phase metallicity play an important role in driving variations in radiation field hardness, as indicated by \NeIII~15.56\m/\NeII~12.81\m, for regions powered by star formation.  Conversely, the radiation hardness for galaxy nuclei powered by accretion around a massive black hole is independent of metal abundance.  Furthermore, for metal-rich environments AGN are distinguishable from star-forming regions by significantly larger \NeIII~15.56\m/\NeII~12.81\m\ ratios.  Finally, \FeII~25.99\m/\NeII~12.81\m\ versus \SiII~34.82\m/\SIII~33.48\m\ also provides an empirical method for discerning AGN from normal star-forming sources.  However, similar to \NeIII~15.56\m/\NeII~12.81\m, these mid-infrared line ratios lose their AGN/star-formation diagnostic powers for very low metallicity star-forming systems with hard radiation fields.
\end {abstract}
 
 
\section {Introduction}
\label{sec:intro}

The {\it Spitzer Space Telescope} has pushed extragalactic infrared spectroscopy into new territory.  {\it Spitzer} observations of silicate emission and absorption in Type~1 and 2 Seyferts, respectively, have provided strong support for the unification model for Active Galactic Nucleus (AGN) galaxies (Sturm et al. 2005; Siebenmorgen et al. 2005; Hao et al. 2005).  Polycyclic Aromatic Hydrocarbons (PAHs) can now be detected and used as redshift indicators for dusty systems at high redshifts (Yan et al. 2005; Weedman et al. 2006a; Teplitz et al. 2007; Valiante et al. 2007).  Spectroscopy from {\it Spitzer} has led to the improvement of mid-infrared diagnostics that characterize and distinguish between ULIRGs, AGN, and star-forming galaxies (Sturm et al. 2006; Weedman et al. 2006b; Dale et al. 2006; O'Halloran, Satyapal, \& Dudik 2006; Brandl et al. 2006; Spoon et al. 2007; Armus et al. 2007; Dudik et al. 2007; Hunter \& Kaufman 2007; Farrah et al. 2007; Sajina et al. 2007, 2008).  Galaxies of low metal abundances exhibit interesting mid-infrared spectral properties as observed by {\it Spitzer}, properties that are attributed to some combination of the inhibition of PAH formation in regions of low metallicity and PAH destruction due to the hard radiation fields present in such systems (e.g., Engelbracht et al. 2005; Cannon et al. 2006; Wu et al. 2006; Engelbracht et al. 2008a; Gordon et al. 2008).  H$_2$ lines have been used to trace shocks and the excitation temperatures and masses of warm molecular hydrogen in galaxies (Devost et al. 2004; Higdon et al. 2006; Appleton et al. 2006; Johnstone et al. 2007; Brunner et al. 2008).  Another interesting result from {\it Spitzer} is that many elliptical galaxies show unusual mid-infrared spectra (Kaneda et al. 2005; Bregman et al. 2006; Kaneda et al. 2008) that may be related to the presence of X-ray emission from low luminosity AGN.  Smith et al. (2007a) suggest that the quiescent environments within ellipticals offer favorable conditions in which to observe these unusual spectra, since the mid-infrared spectra are not dominated by the effects of star formation typically seen in spiral and irregular galaxies.

The {\it Spitzer} Infrared Nearby Galaxies Survey, or SINGS, legacy project obtained a comprehensive spectral database for a large sample of nearby galaxies, and there have been several studies published based on SINGS spectroscopy.  Smith et al. (2004) present first results from science validation galaxy NGC~7331, including one of the first reports of the new PAH complex near 17\m.  Engelbracht et al. (2006), Bendo et al. (2006), and Cannon et al. (2006) explore the strength of the PAH features in different regions of NGC~3034 (M~82), NGC~4594 (M~104), and NGC~1705, respectively.  Dale et al. (2006) use SINGS high resolution spectroscopy for 50 nuclear and 26 extranuclear regions to quantify the power of mid-infrared line diagnostics for distinguishing between star-forming and AGN environments.  Roussel et al. (2006) study silicate, PAH, and H$_2$ emission in the opaque nascent starburst NGC~1377.  Smith et al. (2007a) quantify the energy budget and variations of PAH emission for 59 SINGS nuclei utilizing 5--38\m\ low resolution spectra.  Roussel et al. (2007) study the physical conditions within 57 SINGS nuclei using low and high resolution spectroscopy on H$_2$ lines and \SiII34.82\m.  Engelbracht et al. (2008b) use 20--95\m\ low-resolution IRS~$+$~MIPS SED spectra for 56 SINGS galaxies to compute dust temperatures and the wavelengths at which the mid/far-infrared significantly deviates from a blackbody.  Buckalew et al. (2008) study the impact of ionizing radiation from stellar clusters on the mid-infrared spectra from SINGS extranuclear targets.  Finally, Bot et al. (2008) analyze low resolution Ne, Ar, and S spectral lines from 45 SINGS nuclei to determine the feasibility of constraining elemental abundances using infrared spectroscopy.  

Though there are numerous SINGS papers involving infrared spectroscopy, none has provided the complete set of aperture-matched high resolution spectroscopy data.  Here we present the full compilation of the SINGS high resolution spectral dataset for 88 extranuclear regions and the 67 nuclei with high resolution spectral data.  The data from separate 10--19\m\ and 19--37\m\ instrument modules are extracted from matching 23\arcsec$\times$15\arcsec\ apertures (or 31\arcsec$\times$27\arcsec\ apertures in the case of nine nuclei with extended circumnuclear emission).  The results presented here focus on the applications and interpretations of mid-infrared line ratios.  Compared to previous mid-infrared spectral surveys carried out with the {\it Infrared Space Observatory} (e.g., Thornley et al. 2000; Sturm et al. 2002; Mart\'in-Hern\'andez et al. 2002; Rigopoulou et al. 2002; Vermeij et al. 2002; Verma et al. 2003), {\it Spitzer}/SINGS offers superior sensitivity, and that is the main advantage exploited in this work.  Section~2 describes the sample, Section~3 presents the data, and Section~4 provides the results from this study.  Section~5 summarizes our findings.

\section {The Sample}
\label{sec:sample}

The typical distance for SINGS galaxies is $\sim$10~Mpc, and the distances range from the Local Group out to $\sim$25~Mpc.  The sample spans a wide range of environments and galaxies: low-metallicity dwarfs; quiescent ellipticals; dusty grand design spirals; Seyferts, LINERs, and normal galaxies with starbursting nuclei; and systems within the Local and M~81 groups (see Kennicutt et al. 2003).  However, no SINGS galaxies are exceptionally bright ($L_{\rm FIR}<10^{11}~L_\odot$ in all cases) and none harbors an AGN that dominates the overall luminosity.  

The AGN/star-forming nuclear classifications adopted here are based on the optical spectroscopy described in Moustakas et al. (2009).  In that work three sets of spectroscopic apertures centered on the galaxies are used that cover varying fractions of the integrated $B$ band light: nuclear $\sim$2.5\arcsec$\times$2.5\arcsec\ observations (covering between 0.2 and 1\% of the optical emission), circumnuclear 20\arcsec$\times$20\arcsec\ observations (3--40\%), and $\sim$1\arcmin-wide long-slit drift scan observations (30--100\%).  Though no SINGS AGN dominate the integrated light of the host galaxy---less than 1\% of the total optical emission stems from the central central 2.5\arcsec$\times$2.5\arcsec---the AGN can be energetically significant on the smaller scales studied here.  For example, only 14\% of the nuclei classified as AGN on 2.5\arcsec$\times$2.5\arcsec\ scales switch to the star-forming category when the classifications are carried out on 20\arcsec$\times$20\arcsec\ scales.  More details on the nuclear classifications of SINGS galaxies can be found in Moustakas et al. (2009).

Though the full SINGS sample comprises 75 galaxies, high resolution infrared spectroscopy was obtained for only 67 nuclei.  The nuclei for Holmberg~I, Holmberg~II, M~81~Dwarf~A, M~81~Dwarf~B, DDO~154, IC~2574, and NGC~6822 were deemed too faint in the mid-infrared to merit high resolution spectral observations.  In addition, the nucleus of NGC~3034 (M~82) was reserved by a Guaranteed Time Observer spectroscopy program (Beir{\~a}o et al. 2008).
  
There are a total of 88 extranuclear sources targeted for spectroscopy in the SINGS project, and results from all of them are presented here.  Thirty-nine of these targets were selected based on optical criteria (Kennicutt et al. 2003), while the remaining 49 targets were selected to span a range in infrared properties.  The optically-selected OB/\HII\ regions cover a large range of metallicity, ranging from one-tenth solar to a few times solar, extinction-corrected ionizing luminosity (10$^{49}-10^{52}$~photons~s$^{-1}$), extinction ($A_V \lesssim 4$~mag), radiation field intensity (100-fold range), ionizing stellar temperature ($T_{\rm eff}=35-55$~kK), and local H$_2$/\HI\ ratio as inferred from CO ($<$0.1 to $>$10).  The infrared-selected extranuclear targets were chosen to span a range in $f_\nu$(8.0\m)/$f_\nu$(24\m) and $f$(\hal)/$f_\nu$(8.0\m).  These selection criteria ensured that our observations covered the full range of physical conditions and spectral characteristics found in (bright) infrared sources in nearby galaxies.

\section {Observations and Data Processing}
\label{sec:IRobservations}

\begin{figure}
 \plotone{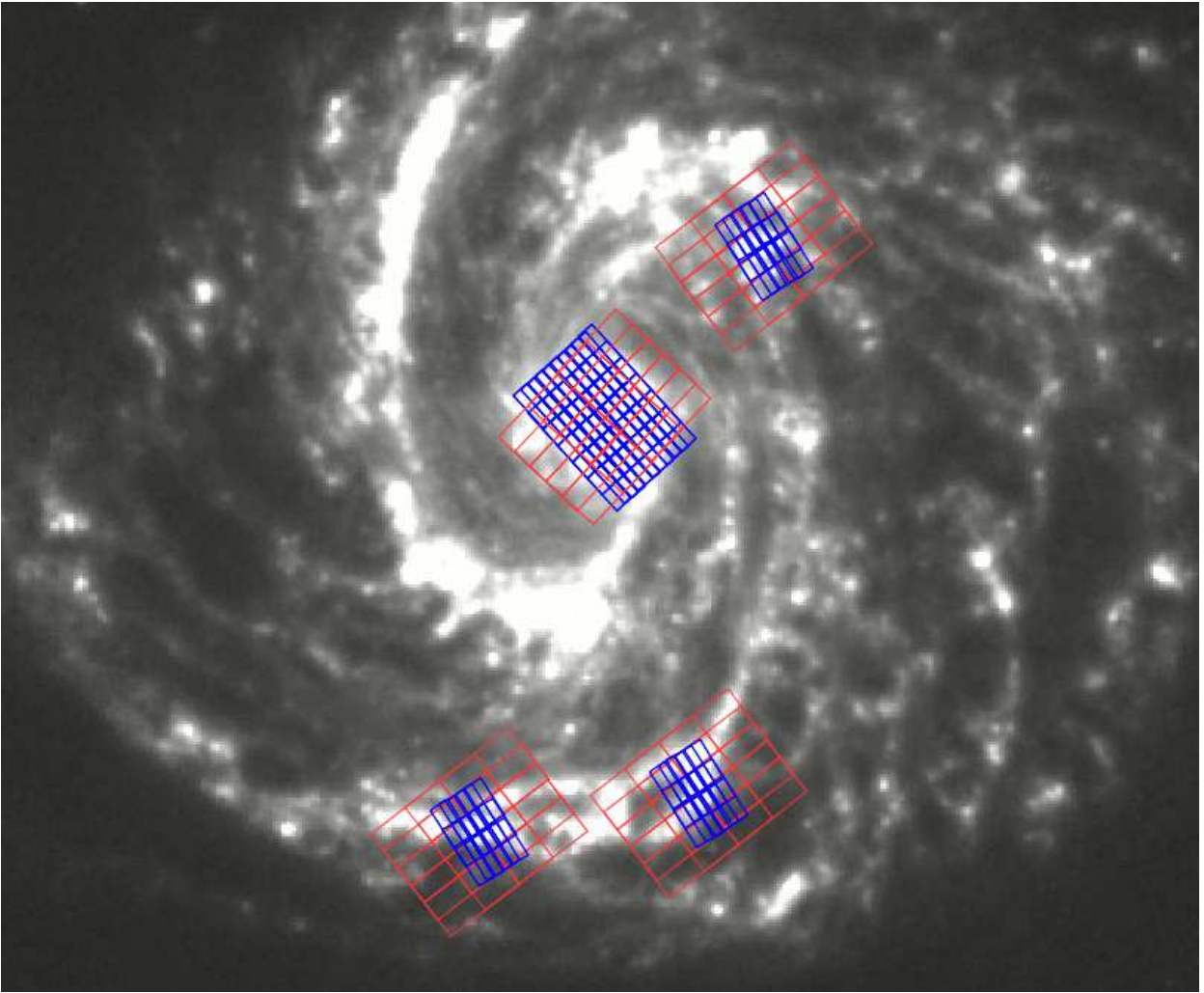}
 \caption{Examples of IRS high resolution mapping footprints overlaid onto an 8\m\ image of NGC~4321.  Short-High mapping is shown in blue and Long-High mapping is portrayed in red.  Most Short-High and Long-High SINGS maps are carried out over a 3x5 pointing grid, though NGC~4321 is one of nine galaxies where an expanded 6x10 grid for Short-High was carried out on the nucleus to better encompass extended circumnuclear emission.}
 \label{fig:overlays}
\end{figure}

High-resolution spectroscopy ($R\sim600$) was obtained in the Short-High (10--19\m) and Long-High (19--37\m) modules (Houck et al. 2004).  Figure~\ref{fig:overlays} displays examples of the general scheme followed for high resolution spectral observations.  High resolution spectroscopy targets were generally mapped with a 3$\times$5 grid in both the Short-High and Long-High modules, utilizing half-slit width and half-slit length steps.  This observational scheme allows for significant data redundancy at each sky position, an important factor if large numbers of bad pixels need to be masked during the construction of the cubes described below.  Owing to the different angular sizes subtended by the instruments, the resulting maps are approximately 45\arcsec$\times$33\arcsec\ in Long-High and 23\arcsec$\times$15\arcsec\ in Short-High.  This mapping scheme made it possible to measure line fluxes with common effective aperture sizes, despite the four-fold change in spatial resolution over the spectral range of our observations.  Slightly larger Short-High nuclear maps were obtained for a subset of nine sources with extended circumnuclear star formation: NGC~1097, NGC~1482, NGC~1512, NGC~1705, NGC~3351, NGC~4321, NGC~4536, NGC~6946, and NGC~7552.  The nine expanded Short-High nuclear maps utilize a 6$\times$10 observational pointing grid and thus span 40\arcsec$\times$28\arcsec.  All integrations are 60~s per pointing.  The effective integrations are longer since each location was observed 2--4 times.  

The individual data files for a given spectral map were assembled into spectral cubes using the software CUBISM (Smith et al. 2007b).  The cube input data were pre-processed using the latest versions of the {\it Spitzer Science Center} pipeline: S15.3 for Short-High and S17.2 for Long-High.  The S17.2 Long-High data include improved corrections for the order-to-order continuum tilts that are present in the S15.3 Long-High data.  As no sky observations were obtained at high spectral resolution, the high resolution spectra are not sky-subtracted.  Various processing steps within CUBISM are described by Smith et al. (2007b).  One-dimensional spectra were extracted from most of the Short-High and Long-High data cubes using matching $\sim$23\arcsec$\times$15\arcsec\ apertures.  For the nine galaxies with extended Short-High maps described above, the extraction apertures are $\sim$31\arcsec$\times$27\arcsec.  In terms of physical scales, the projected areal coverage of the extractions spans the range 0.002--11~kpc$^2$ with a sample median of 0.48~kpc$^2$.\footnote{The distances in the SINGS sample span a factor of $\sim$50 and the high resolution spectral maps are relatively small, so it is difficult to carry out interesting, sample-wide analyses of the high resolution spectral maps that utilize apertures matched in physical size.}  The extraction centers and projected areas are listed in Table~\ref{tab:fluxes1}.  Figure~\ref{fig:spectra1} shows all the high resolution spectra for the 67 nuclear and 88 extranuclear regions observed.\footnote{SINGS data, including these spectra, are found at http://ssc.spitzer.caltech.edu/legacy/singshistory.html.}

Emission line fluxes are derived from continuum-subtracted Gaussian fits to the lines and first or second order polynomial fits to the continua; second order continuum fits are used for cases like \NeII 12.81\m, which is blended with a PAH emission feature.  Based on extrapolations from the 19-31\m\ continuum, the pipeline corrections for the order-to-order continuum tilts mentioned above appear to be overestimated for the longest two wavelength ($\lambda\gtrsim31$\m) orders in S17.2 data.  Hence, our emission line flux estimates for \SIII~33.48\m\ and \SiII~34.82\m\ are averaged from the S15.3 and S17.2 pipelines.  These S15.3/S17.2 average line fluxes deviate by $\sim$8\% and 1.5\%, respectively, from the individual S15.3 and S17.3 pipeline extractions.  The emission line fluxes are listed in Tables~\ref{tab:fluxes1} and \ref{tab:fluxes2} along with 3$\sigma$ upper limits for nondetections in several key lines.  Upper limits are based on the spectral resolution $\Delta \lambda$ of the high resolution spectrometer and the root-mean-square variations in the flux density of the local continuum: 
\be
f(3\sigma)=3 f_\lambda({\rm r.m.s.}) \Delta \lambda=3 f_\lambda({\rm r.m.s.}) \lambda/600.
\ee  
It is occasionally difficult to assess from the one-dimensional spectra alone whether or not a faint line is detected.  The fully-processed two-dimensional spectra were thus inspected within CUBISM to verify each detection and each non-detection.

Several cross-checks on the flux calibration were made between data from the various IRS modules and {\it Spitzer} imaging data at 8.0 and 24\m\ and {\it Infrared Space Observatory} imaging data at 6.75 and 15\m.  The cross-checks were carried out by comparing the broadband imaging flux densities to the corresponding simulated values obtained by integrating under the spectra weighted according to the imaging filter bandpass profiles.  The absolute flux calibration uncertainty for the spectral data is estimated to be 25\% (1$\sigma$); the uncertainty in line flux ratios is $\sim$10\% (see also the SINGS Fifth Data Release delivery document at http://ssc.spitzer.caltech.edu/legacy/singshistory.html).

\section {Results}

\subsection{Nuclear and Extra-Nuclear Mid-Infrared Spectra}
\label{sec:spectra}
Figure~\ref{fig:spectra1} displays all 155 10--37\m\ high resolution spectra for the SINGS sample.  The continuum levels observed by the separate Short-High (10--19\m) and Long-High (19--37\m) modules were not adjusted to be equal, but they do roughly match each other in most cases.  The agreement between Short-High and Long-High at $\sim$19.3\m\ is within 0.5~$\cdot 10^{-6}$~W~m$^{-2}$~sr$^{-1}$ for 95\% of the cases.  Expressed relative to the continuum levels, $\sim$95\% of the Short-High and Long-High spectra match to within 33\%; only 5\% of the Short-High and Long-High data have significant amplitude mismatches relative to the continuum level, and in each of these cases the large relative discrepancy is due to a weak continuum.  For many of the faintest galaxies, the Long-High continuum slopes downwards for longer wavelengths (e.g., NGC~1404).  In these cases we are seeing the effects of foreground zodiacal emission, not emission from the target galaxy, since we have not sky-subtracted the high resolution spectra.

The strongest features in these spectra are typically the 11.3, 12.7, and 17\m\ PAH complexes (see Smith et al. 2007a) and the \NeII 12.81\m, \NeIII 15.56\m, \SIII 18.71\m, \FeII 25.99\m, \SIII 33.48\m, and \SiII 34.82\m\ emission lines.  Weaker or more variable features include the higher ionization forbidden lines \SIV10.51\m, \NeV14.32\m, and \OIV25.89\m, and the molecular hydrogen lines at 12.28, 17.04, and 28.22\m.  The $\nu=0-0$ S(0), S(1), and S(2) molecular hydrogen lines from SINGS were studied in detail by Roussel et al. (2007), and one of the main results from that work is the discovery of a tight link between the H$_2$ line intensities and the PAH emission, over a large range in radiation field intensities (see also Rigopoulou et al. 2002).  

Figure~\ref{fig:blowup} shows many of the above mid-infrared emission features over smaller wavelength ranges for the bright nucleus of NGC~5194 (M~51).  Much of the infrared fine structure line emission from the SINGS sample derives from nebular lines from \HII\ regions.  However, there are several exceptions.  \SiII34.82\m, for example, comes from a wider variety of environments, including both ionized gas and warm atomic gas such as photodissociation regions and X-ray dominated regions (e.g., Hollenbach \& Tielens 1999).  The spatial variability of \SiII34.82\m\ will be explored in a future paper using the much more spatially extensive SINGS low-resolution spectral maps.  Iron also has an ionization potential smaller than 13.6~eV, so similar to \SiII34.82\m\ the \FeII25.99\m\ line can be strong outside of \HII\ regions, and in fact is frequently used as an indicator of shocks from supernova activity (e.g., Alonso-Herrero et al. 2003) or AGN via X-ray photoionization processes (e.g., Knop et al. 2001).  In addition, Ne$^{3+}$ has such a high ionization potential ($\sim$97~eV) that \NeV~14.32\m\ in galactic nuclei is only observed in AGN (Satyapal et al. 2007); $\sim$55~eV are necessary to create O$^{3+}$, so \OIV~25.89\m\ is also frequently used as a tracer of AGN (see Lutz et al. 1998; Genzel et al. 1998).  The ionization potentials needed to create the species for each fine structure line are provided in Table~\ref{tab:fluxes2}.  The unusual mid-infrared spectrum of the NGC~1377 nucleus is described as a heavily dust-obscured nascent starburst in Roussel et al. (2007).  Comparing the nuclear and extranuclear spectra, there is one broad generalization that immediately catches the eye:  the proportion of spectra with large line-to-continuum ratios is higher for the extranuclear regions than for the nuclei.  This general result is expected, though, since the extranuclear targets are primarily actively forming stars, whereas the SINGS nuclei have more luminous continuum levels and are far more diverse, spanning a larger range in star formation activity and hosting varying levels of AGN activity (\S~\ref{sec:sample}).  The same tendency for nuclei to have lower line-to-continuum ratios than disk \HII\ regions is also seen in optical spectra (e.g., Kennicutt et al. 1989).

\subsection{Radiation Field Hardness and Oxygen Abundance}
\label{sec:hardness}
The hardness of the interstellar radiation field is influenced by several factors, including the age of the stellar population producing the bulk of the light, the characteristic metallicity of the environment, and the properties of the AGN's accretion disk when a massive central black hole is present.  In this section the radiation field hardness is explored as a function of metal abundance.  The oxygen abundances utilized in this work are from Moustakas et al. (2009), who have coupled new optical spectroscopic data with archival optical emission line measurements from 570 SINGS \HII\ regions) to derive abundances and radial abundance gradients for most SINGS galaxies.  The abundances used for the extranuclear sources are computed from their (de-projected) radial distances and parent galaxy radial abundance gradients.  The abundances used for the nuclear sources are the ``characteristic'' abundances at a radial position of 40\% of the optical radius, 0.4$R_{25}$, a definition which avoids issues related to AGN contamination.  The characteristic abundance is representative of the mean, optical-luminosity-weighted oxygen abundance of each galaxy.  The strong-line calibration of Kobulnicky \& Kewley (2004) is adopted for the abundances in this work.  

Several emission line and PAH diagnostics were presented in Dale et al. (2006) based on a subset (about half) of the SINGS high resolution spectral database.  Figure~\ref{fig:Z} shows another emission line ratio diagram, using the entire SINGS high resolution spectral database: \NeIII 15.56\m/\NeII 12.81\m\ ratio as a function of the gas-phase oxygen abundance (see Verma et al. 2003 for a similar analysis using {\it ISO} data).  The \NeIII~15.56\m/\NeII~12.81\m\ ratio is considered an indicator of the hardness of the interstellar radiation field (e.g., Thornley et al. 2000; Madden et al. 2006; Beir{\~a}o et al. 2008; Gordon et al. 2008).  Perhaps the most important result from Figure~\ref{fig:Z} is that star-forming nuclear and extranuclear targets show the same general trend: harder radiation fields for lower metal abundance environments, consistent with prior optical studies (e.g., Searle 1971).  This trend spans a factor of 150 in \NeIII 15.56\m/\NeII 12.81\m\ and a factor of 10 in metallicity.  Theoretical predictions from photoionization models are superposed onto the data displayed in the righthand panel of Figure~\ref{fig:Z}.  The models assume an 8~Myr continuous Starburst99 (Leitherer et al. 1999) star formation scenario for an electron density of 350~cm$^{-3}$.  In order from lowest to highest, the models overlaid correspond to $q=Uc=2\cdot10^7$, $4\cdot10^7$, $8\cdot10^7$, $1.5\cdot10^8$, and $3\cdot10^8$~cm~s$^{-1}$, where the ionization parameter $U$ is defined as the ionizing photon flux per unit area normalized to the gas number density:
\be
U=Q_{\rm H}/(4\pi r^2nc),
\ee
with $Q_{\rm H}$ representing the number of ionizing photons per second, $n$ the gas number density, and $r$ the separation between illumination source and cloud.  This range of photoionization models does a fair job of bracketing the observed trend in radiation hardness and metallicity for star-forming regions.  The metallicity of galaxies with AGN, on the other hand, empirically plays little role in the radiation field hardness.  Rather, the radiation field hardness for AGN systems likely depends more on the temperature distribution within the X-ray-emitting accretion disk, contributions from non-thermal processes such as synchrotron radiation, and the AGN/star formation luminosity ratio within our apertures.  

Star-forming regions show a larger dynamic range than the LINER/Seyfert regions in radiation field hardness (a factor of 150 versus a factor of 30 in \NeIII~15.56\m/\NeII~12.81\m); the radiation field hardness in the two populations reaches down to a similar lower bound, but ranges to substantially higher values for the lowest-metallicity star-forming regions.  In a {\it Spitzer}/IRS study of eight classical AGN, Weedman et al. (2005) find a similar range in \NeIII~15.56\m/\NeII~12.81\m, with ratios spanning a range from $\sim$0.17 to 1.9.  Assuming that our catchall term ``AGN'' is comprised of both LINERs and Seyferts, it is somewhat surprising that star-forming regions show a much larger hardness range than AGN since LINERs are known to have ionization parameters an order of magnitude lower than Seyferts (e.g., Ho, Filippenko, \& Sargent 2003 and references therein).  In addition to the lack of dependence of radiation hardness on metallicity, presumably SINGS AGN do not exhibit extremely large \NeIII 15.56\m/\NeII 12.81\m\ ratios partly due to a larger fraction of neon residing in ionization states higher than Ne$^{2+}$.  This interpretation is anecdotally supported by the detection rate of \NeV14.32\m\ in the SINGS sample: \NeV14.32\m\ is detected in four of 35 AGN but in none of the 120 star-forming targets.  For comparison Weedman et al. (2005) detect \NeV14.32\m\ in four of eight AGN studied.

Though the lowest-metallicity star-forming sources exhibit the hardest radiation fields, and despite the fact that a substantial portion of the neon in AGN may reside in ionization states higher than Ne$^{2+}$, overall the AGN show larger average \NeIII 15.56\m/\NeII 12.81\m\ ratios (see also Madden et al. 2006; Kaneda et al 2008).  Statistically, the extranuclear and star-forming nuclear regions exhibit a median and dispersion in \NeIII 15.56\m/\NeII 12.81\m\ of 0.28 and 0.49~dex, respectively, whereas the AGN show a median of 0.53 and a spread of 0.38~dex in the ionized neon ratio.  Furthermore, note that for super-solar metallicity systems ($12+\log({\rm O/H})\gtrsim9.0$ for the Kobulnicky \& Kewley (2004) strong-line calibration), the average hardness of the radiation field is conspicuously more elevated for AGN than for star-forming regions.  For the region defined by $12+\log({\rm O/H})\gtrsim9.0$, AGN cluster mainly between 0.04 and 2 in the ratio \NeIII 15.56\m/\NeII 12.81\m\ whereas the star-forming regions group between 0.006 and 0.04.  The ability to distinguish between AGN and star formation using a \NeIII/\NeII\ hardness indicator, at least for metal rich environments, is consistent with theoretical expectations (e.g., Figure~6g of Spinoglio \& Malkan 1992; \S~3.2.3 of Voit 1992).

\subsection{OIV/NeII and PAH Strength}

Studies of AGN, starbursting galaxies, and ultraluminous infrared galaxies with the {\it Infrared Space Observatory} showed that ratios of high ionization-to-low ionization lines could separate AGN-dominated sytems from sytems powered by star formation (e.g, Genzel et al. 1998; Sturm et al. 2002; Peeters, Spoon, \& Tielens 2004).  When such ratios are plotted as a function of PAH strength, the separation is accentuated.  Figure~\ref{fig:genzel} displays this type of diagram for SINGS targets based on {\it Spitzer} spectroscopy.  The SINGS sample does not contain luminous infrared galaxies ($L_{\rm FIR}\gtrsim10^{11}~L_\odot$) or ultraluminous infrared galaxies($L_{\rm FIR}\gtrsim10^{12}~L_\odot$), and probes relatively faint dwarf galaxies and individual \HII\ regions.  Nevertheless, these mid-infrared diagnostics for the relatively lower-luminosity SINGS sample show a separation between AGN and star-forming regions similar to that seen for higher luminosity targets.

\subsection{FeII/NeII and SiII/SIII}
Figure~\ref{fig:line_ratios} shows \FeII~25.99\m/\NeII~12.81\m\ as a function of \SiII~34.82\m/\SIII~33.48\m.  This line ratio combination shows a strong correlation, and more strikingly, a clear separation between star-forming and AGN regions.  To understand the correlation as well as the separation, it is instructive to review the similarities in the two line ratios.  The numerators utilize Si and Fe, refractory elements that provide two of the main building blocks for dust grains (Draine 2003).  Both numerators involve species with similar ionization potentials ($\sim$8~eV) that are below one Rydberg, whereas the denominators involve species with ionization potentials of 22--23~eV; Fe$^+$ and Si$^+$ can emit from outside of \HII\ regions while Ne$^+$ and S$^{2+}$ emit from within \HII\ regions.

One possibility for the AGN/star-forming separation in Figure~\ref{fig:line_ratios} involves the extent to which Si and Fe are depleted onto dust grains.  Perhaps the hard radiation fields concomitant with AGN provide enough sufficiently energetic photons such that photodesorption returns a large number of Si and Fe atoms from the surfaces of dust grains back to the gas phase of the interstellar medium.  In addition, AGN are able to heat dust to very high temperatures out to large radii, potentially leading to thermal sublimation of interstellar grains, or it's possible that the dense environment leads to enhanced grain erosion through sputtering or even shattering (see Draine 2003 for a review).  Thus the liberation of dust grain constituents, as indicated by gas phase Fe and Si lines, may be enhanced for active galaxies.  A difficulty with this interpretation is that Si and Fe can also be returned to the gas phase via shocks associated with individual supernovae or with starburst-driven superwinds, or even mergers (Forbes \& Ward 1993; Alonso-Herrero et al. 2003; O'Halloran, Satyapal, \& Dudik 2006; Armus et al. 2006; Armus et al. 2007), so it's unclear whether the level of Si and Fe depletion is the dominant factor for the AGN/star-forming separation observed in Figure~\ref{fig:line_ratios}.

Another interpretation for the observed separation in Figure~\ref{fig:line_ratios} relies on X-ray photoionization processes associated with AGN (e.g., Knop et al. 2001).  According to predictions from X-ray dissociation models (e.g., Maloney et al. 1996), very energetic ($\gtrsim$keV) photons are able to travel large distances through the interstellar medium due to their small cross-sections to absorption.  Once these energetic photons finally are absorbed and deposit their energy into the interstellar medium, a large, low ionization volume is produced via a slew of secondary ionizations.   Along with several prominent ultraviolet and optical emission lines (e.g., Ly$\alpha$, \OI~6300\AA, \CI~9823,9850\AA), a few infrared emission lines are predicted to be particularly bright from X-ray dissociation regions, including \OI~63.18\m, \SiII~34.82\m, \FeII~25.99\m, \OI~145.5, and \CII~157.7\m.  Observations support this scenario of bright XDR lines for AGN-like environments (Veilleux \& Osterbrock 1987; Armus, Heckman, \& Miley 1989; Veilleux 1991; Spinoglio \& Malkan 1992; Osterbrock 1993; Dale et al. 2004).  Compared to sites of star formation, a smaller fraction of the energy budget in XDRs results in nebular line emission such as \NeII~12.81\m\ and \SIII~33.48\m.  The observed range of \FeII~25.99\m/\NeII~12.81\m\ for AGN can be recovered for some X-ray dissociation models (e.g., Maloney 1999; P. Maloney, priv. comm.), but the unknown Si and Fe depletion factors and ionization parameters complicate any model-based conclusions.

A third explanation for the unusually high proportion of \SiII~34.82\m, and \FeII~25.99\m\ emission for AGN environments is based on interstellar density.  Theoretical models show that the ratios \SiII(PDR)/\SiII(\HII), \FeII(PDR)/\FeII(\HII), \SiII(XDR)/\SiII(PDR), and \FeII(XDR)/\FeII(PDR) increase with increasing density (Kaufman, Wolfire, \& Hollenbach 2006; Meijerink \& Spaans 2005).  So if the mid-infrared line-emitting gas in AGN is typically more dense than found in starbursts and normal galaxies, this could lead to the observed enhancement in \SiII~34.82\m\ and \FeII~25.99\m\ emission.  As will be shown in Section~\ref{sec:density}, there is no evidence that the \HII\ regions within apertures centered on AGN are significantly more dense compared to sites dominated by star formation, even when the extraction apertures are shrunk by a factor of three to four in area.  It is conceivable that the bulk of the mid-infrared emitting regions associated with these AGN may be far more compact than our apertures, but we can at least report that the use of apertures three to four times smaller results in no significant change in Figure~\ref{fig:line_ratios}.

The above interpretations focus on the emission from Si and Fe, but it is also possible that the trends observed in Figure~\ref{fig:line_ratios} are governed by the line emission mechanisms responsible for the normalizations in these line ratios.  For example, the separation between AGN and star-forming regions holds when either of the two normalizations in Figure~\ref{fig:line_ratios} are replaced by a third nebular line with a similar $\sim$23~eV ionization potential: \SIII~18.71\m.  Furthermore, the distribution in Figure~\ref{fig:line_ratios} hints that extranuclear regions, presumably more representative of ``pure'' \HII\ regions, slightly separate from the star-forming nuclei, further supporting the notion that nebular lines drive the correlation.  In addition, many AGN show a higher fraction of elements residing in highly ionized states (\S~\ref{sec:hardness}); relatively weak low-ionization emission in AGN can drive these ratios to higher values than seen in star-forming regions.  

Consistent with this picture is the finding by O'Halloran et al. (2006) of relatively large ratios of \FeII~25.99\m/\NeII~12.81\m\ for very low metal abundance systems.  The hard radiation fields for such systems (\S~\ref{sec:hardness}) expend comparatively more of their radiative energy on producing higher ionization lines such as \NeIII~15.56\m\ and less on relatively lower ionization lines like \NeII~12.81\m.  It is also possible that low-metallicity regions have lower Fe depletion rates onto grains.  In this narrow sense low metallicity systems are observationally similar to AGN, and show relatively large ratios of \FeII~25.99\m/\NeII~12.81\m.  Note that the SINGS targets with the lowest metallicities do not appear in Figures~\ref{fig:Z} and \ref{fig:line_ratios} since they were either not observed with the IRS high resolution spectrometer (e.g., M81~Dwarf~A, M81~Dwarf~B, IC~2574, DDO~154, NGC~6822, Holmberg~I, and Holmberg~II) or were not detected in these lines (e.g., Holmberg~IX, DDO~053, DDO~165, NGC~5408, and NGC~5474).

\subsection{Interstellar Density}
\label{sec:density}
Figure~\ref{fig:density} shows the distribution of nebular electron densities as determined by the line ratio \SIII 18.71\m/\SIII 33.48\m (see the top axis of the figure; Mart\'in-Hern\'andez et al. 2002).  The average density is calculated using electron collision strengths from Tayal \& Gupta (1999).  The effects of differential extinction at these mid-infrared wavelengths have not been included in this calculation, but the internal extinction for SINGS galaxies at these wavelengths is negligible (Dale et al. 2006; Prescott et al. 2007; Smith et al. 2007a), except for NGC~1377 which does not appear in Figure~\ref{fig:density} since neither sulfur line is securely detected (Roussel et al. 2006).  Though the SINGS sample exhibits a wide range in interstellar electron densities, the {\it average} values for the three different environments are similar.  The median densities for star-forming nuclear, extranuclear, and AGN regions are respectively 270, 380, and 280~cm$^{-3}$.  The lack of significant differences in average density between different environments could be partly attributable to the $\sim$kiloparsec scales sampled by our apertures---each aperture presumably captures emission from both \HII\ and photodissociation regions, in addition to the emission from the accretion disks for some apertures (e.g., Dudik et al. 2007).  To investigate this possibility we have re-extracted the spectra over $\sim$11\arcsec$\times$9\arcsec\ apertures.  Over these smaller apertures, which cover $\sim$29\% of the area spanned by the $\sim$23\arcsec$\times$15\arcsec\ apertures, the median inferred densities are 60-70\% larger, or 460, 600, and 460~cm$^{-3}$ for star-forming nuclear, extranuclear, and AGN regions, respectively.  
Due to the diffraction limitations of the telescope we cannot meaningfully take this type of comparison to smaller spatial scales, but on the relatively small scales probed here the different environments retain their density similarities.

\section {Summary}

The SINGS sample of 75 galaxies and 88 extranuclear regions covers a broad swath of the extragalactic environments found in the local universe.  Faint low metallicity systems, bright starbursting galaxies, quiescent and star-forming ellipticals, low luminosity AGN, and ``normal'' spirals are all represented in the SINGS sample.  Important legacy aspects to the SINGS project are the comprehensive, uniform databases of {\it Spitzer} imaging and low- and high-resolution spectroscopy, along with the extensive ancillary ultraviolet, optical, near-infrared, submillimeter, and radio datasets.  This contribution focuses on the SINGS high resolution mid-infrared spectroscopy.  Short-High (10--19\m) and Long-High (19--37\m) spectra from matching apertures are presented for the full SINGS sample of nuclear and extranuclear regions.  The majority of apertures are $\sim$23\arcsec$\times$15\arcsec, though 31\arcsec$\times$27\arcsec\ apertures are utilized for nine nuclei with extended circumnuclear emission for which we obtained larger Short-High maps.  The fluxes for nine atomic forbidden and three molecular hydrogen mid-infrared emission lines are provided in addition to upper limits in several key lines for infrared-faint targets.  

Three mid-infrared line diagnostics present opportunities to distinguish between star formation and accretion-powered systems.  Using a mid-infrared line diagnostic that involves the radiation field hardness indicator \NeIII 15.56\m/\NeII 12.81\m, it is shown that for regions with super-solar abundances of oxygen the \NeIII 15.56\m/\NeII 12.81\m\ ratio is elevated for AGN.  Moreover, and more importantly, metallicity is associated with radiation field hardness for regions powered by star formation.  The observed trend for star-forming targets matches photoionization models using standard inputs.  The same dependence on metallicity does not hold true for active galactic nuclei, systems for which the radiation field hardness is more likely driven by the properties of the resident accretion disk. 

PAH equivalent widths and ratios of mid-infrared high ionization-to-low ionization lines both help to discriminate between AGN and star-forming nuclei, a technique first pioneered by {\it Infrared Space Observatory} observations of AGN, starbursting, and ultraluminous infrared galaxies.  Lower luminosity SINGS AGN, star-forming nuclei, and star-forming extra-nuclear regions follow the same general trend first seen with {\it ISO}.

In a third mid-infrared line diagnostic, the ratios \FeII~25.99\m/\NeII~12.81\m\ and \SiII~34.82\m/\SIII~33.48\m\ are smaller for normal star-forming environments than for AGN.  This striking result is presumably due to depressed \NeII~12.81\m\ and \SIII~33.48\m\ emission, and/or enhanced \FeII~25.99\m\ and \SiII~34.82\m\ emission for sites powered by AGN.  The former would be due to increased ionization of the species, and the latter to increased emission from X-ray dissociation regions, or possibly to decreased depletion of iron and silicates onto interstellar dust grains in AGN-powered environments; Fe and Si atoms are perhaps more easily liberated from dust grains into the gas phase in the hot accretion disk environments associated with AGN.  However, the ability to discriminate between AGN and star formation breaks down for environments with very low metallicities.  The hardness of the radiation fields for systems with low metal abundances appears to blur the lines between AGN and low-metallicity star formation, at least in the narrow sense of the line ratios \FeII~25.99\m/\NeII~12.81\m, \SiII~34.82\m/\SIII~33.48\m, and \NeIII 15.56\m/\NeII 12.81\m.  

Finally, the ratio of \SIII~18.71\m/\SIII~33.48\m\ indicates a wide range of interstellar electron densities but the average value on $\sim$23\arcsec$\times$15\arcsec\ ($\sim$11\arcsec$\times$9\arcsec) scales is 300--400~cm$^{-3}$ (500--600~cm$^{-3}$) for star-forming nuclear, star-forming extranuclear, and AGN environments.

\acknowledgements 
Rajib Ganguly, Brent Groves, and Phil Maloney graciously helped with comparisons of the data to theoretical models.  We are grateful for the helpful suggestions provided by the referee.  Support for this work, part of the {\it Spitzer Space Telescope} Legacy Science Program, was provided by NASA through Contract Number 1224769 issued by the Jet Propulsion Laboratory, California Institute of Technology under NASA contract 1407.  This research has made use of the NASA/IPAC Extragalactic Database which is operated by JPL/Caltech, under contract with NASA.  This publication makes use of data products from the Two Micron All Sky Survey, which is a joint project of the University of Massachusetts and the Infrared Processing and Analysis Center/California Institute of Technology, funded by the National Aeronautics and Space Administration and the National Science Foundation.

\begin{deluxetable}{lcrrrr}
\rotate
\def\a{\tablenotemark{a}}
\def\b{\tablenotemark{b}}
\def\c{\tablenotemark{c}}
\def\p{$\pm$}
\tabletypesize{\scriptsize}
\tablenum{1}
\label{tab:fluxes1}
\tablecaption{Nuclear and ExtraNuclear Emission Line Fluxes: Molecular Hydrogen}
\tablewidth{0pc}
\tablehead{
\colhead{Target} &
\colhead{$\alpha_0~~~\delta_0$} &
\colhead{Projected} &
\colhead{H$_2$ S(2)} &
\colhead{H$_2$ S(1)} &
\colhead{H$_2$ S(0)} 
\\
\colhead{} &
\colhead{(J2000.0)} &
\colhead{Area} &
\colhead{12.28$\mu$m} &
\colhead{17.04$\mu$m} &
\colhead{28.22$\mu$m} 
\\
\colhead{} & 
\colhead{} & 
\colhead{(kpc$^2$)} & 
\colhead{1682~K} &
\colhead{1015~K} &
\colhead{510~K} 
}
\startdata
NGC~0024     &000956.3 $-$245750& 0.427&\nodata      &  0.82\p 0.42&  0.94\p 0.22\nl
NGC~0337     &005950.0 $-$073441& 3.975&\nodata      &  2.95\p 0.53&  1.86\p 0.49\nl
NGC~0584\a   &013120.7 $-$065204& 3.154&\nodata      &\nodata      &\nodata      \nl
NGC~0628     &013641.7 $+$154701& 0.463&\nodata      &  2.18\p 0.53&  0.91\p 0.27\nl
NGC~0855     &021403.6 $+$275239& 0.721&\nodata      &  2.71\p 1.16&\nodata      \nl
NGC~0925     &022717.1 $+$333445& 0.660&\nodata      &  1.65\p 0.47&  0.63\p 0.14\nl
NGC~1097\a\b &024619.0 $-$301629& 5.732& 19.99\p 1.66& 43.10\p 0.81& 12.72\p 3.76\nl
NGC~1266\a   &031600.7 $-$022538& 7.144& 17.34\p 0.86& 21.76\p 0.71&\nodata      \nl
NGC~1291\a   &031718.6 $-$410629& 0.928&  1.50\p 0.44&  5.01\p 0.62&  1.15\p 0.26\nl
NGC~1316\a   &032241.7 $-$371230& 4.686&  2.84\p 0.54&  6.09\p 0.59&\nodata      \nl
NGC~1377\a   &033639.1 $-$205408& 5.842&  3.53\p 0.92&  9.47\p 0.82&\nodata      \nl
NGC~1404\a   &033851.9 $-$353539& 2.694&\nodata      &\nodata      &\nodata      \nl
NGC~1482\b   &035438.9 $-$203008&10.583& 13.58\p 0.82& 24.49\p 0.84&  5.76\p 1.55\nl
NGC~1512\a\b &040354.2 $-$432055& 2.739&\nodata      &  4.84\p 0.32&  1.47\p 0.16\nl
NGC~1566\a   &042000.4 $-$545617& 3.582&  8.25\p 0.61& 17.01\p 0.61&  3.10\p 0.38\nl
NGC~1705\b   &045413.6 $-$532138& 0.511&\nodata      &  0.29\p 0.13&\nodata      \nl
NGC~2403     &073650.0 $+$653604& 0.082&\nodata      &\nodata      &  0.74\p 0.18\nl
DDO~053      &083407.2 $+$661054& 0.101&\nodata      &\nodata      &\nodata      \nl
NGC~2798     &091722.9 $+$420000& 5.432& 15.10\p 1.41& 31.52\p 0.94&\nodata      \nl
NGC~2841\a   &092202.6 $+$505835& 0.844&  1.32\p 1.29&  1.52\p 0.31&  0.46\p 0.13\nl
NGC~2915     &092611.5 $-$763736& 0.113&\nodata      &\nodata      &\nodata      \nl
NGC~2976     &094715.3 $+$675500& 0.100&\nodata      &  2.57\p 0.63&  1.30\p 0.14\nl
NGC~3049     &095449.6 $+$091617& 4.986&\nodata      &  2.66\p 0.35&  1.42\p 0.58\nl
NGC~3031\a   &095533.2 $+$690355& 0.104&  4.30\p 0.63&  7.17\p 0.38&  0.95\p 0.34\nl
Holmberg~IX  &095732.0 $+$690245& 0.086&\nodata      &\nodata      &\nodata      \nl
NGC~3190\a   &101805.6 $+$214956& 3.797&  3.14\p 0.44&  9.35\p 0.44&  3.04\p 0.21\nl
NGC~3184     &101816.9 $+$412528& 0.671&  1.64\p 0.34&  3.56\p 0.47&  1.40\p 0.27\nl
NGC~3198\a   &101954.9 $+$453259& 1.488&  2.30\p 0.79&  4.85\p 0.84&  2.41\p 0.52\nl
NGC~3265     &103106.7 $+$284748& 4.256&  2.36\p 1.03&  4.36\p 0.40&  1.69\p 0.48\nl
Markarian33  &103231.9 $+$542404& 4.510&  2.03\p 0.60&  3.90\p 0.87&  2.48\p 0.62\nl
NGC~3351\b   &104357.7 $+$114213& 1.712&  5.52\p 0.64& 13.55\p 0.43&  3.81\p 1.05\nl
NGC~3521\a   &110548.6 $-$000209& 0.793&  1.72\p 0.80&  5.06\p 0.85&  2.43\p 0.32\nl
NGC~3621\a   &111816.5 $-$324851& 0.347&  2.66\p 0.69&  6.18\p 0.49&  2.41\p 0.17\nl
NGC~3627\a   &112015.0 $+$125930& 0.759& 21.44\p 0.75& 44.49\p 0.78&  6.02\p 0.71\nl
NGC~3773     &113813.0 $+$120644& 1.125&\nodata      &  1.05\p 0.23&  0.90\p 0.19\nl
NGC~3938\a   &115249.4 $+$440715& 1.650&\nodata      &  1.84\p 0.28&  1.29\p 0.19\nl
NGC~4125\a   &120806.0 $+$651028& 4.156&  1.33\p 0.40&  3.12\p 0.51&  0.62\p 0.10\nl
NGC~4236     &121642.1 $+$692746& 0.157&\nodata      &\nodata      &\nodata      \nl
NGC~4254     &121849.6 $+$142500&10.959&  5.09\p 0.61& 13.94\p 0.79&  2.92\p 0.40\nl
NGC~4321\a\b &122254.9 $+$154921& 4.022&  8.26\p 0.90& 15.35\p 0.37&  5.36\p 0.40\nl
NGC~4450\a   &122829.6 $+$170506& 7.858&  5.68\p 0.80& 15.39\p 0.55&  1.74\p 0.24\nl
NGC~4536\b   &123427.1 $+$021117& 4.091& 11.31\p 0.90& 26.89\p 1.02&  6.85\p 1.66\nl
NGC~4552\a   &123539.8 $+$123323& 2.008&\nodata      &\nodata      &\nodata      \nl
NGC~4559     &123557.7 $+$275736& 0.840&\nodata      &  3.01\p 0.39&  1.89\p 0.18\nl
NGC~4569\a   &123649.8 $+$130946& 3.166& 24.69\p 0.89& 48.48\p 0.62&  5.96\p 0.89\nl
NGC~4579\a   &123743.6 $+$114906& 2.184& 16.58\p 0.82& 25.05\p 0.45&  1.72\p 0.27\nl
NGC~4594\a   &123959.5 $-$113723& 0.694&\nodata      &\nodata      &  0.54\p 0.38\nl
NGC~4625\a   &124152.6 $+$411626& 0.666&  1.09\p 0.38&  2.12\p 0.37&  1.22\p 0.20\nl
NGC~4631     &124207.8 $+$323235& 0.604&  8.35\p 0.95& 16.71\p 0.52& 10.28\p 0.82\nl
NGC~4725\a   &125026.6 $+$253003& 1.223&  2.63\p 0.67&  5.97\p 0.58&  1.79\p 0.21\nl
NGC~4736\a   &125053.1 $+$410713& 0.217& 14.78\p 1.02& 33.21\p 0.99&  4.18\p 0.68\nl
NGC~4826\a   &125643.7 $+$214100& 0.194& 22.86\p 1.79& 51.42\p 0.74& 11.11\p 0.62\nl
 DDO~165     &130624.8 $+$674225& 0.165&\nodata      &\nodata      &\nodata      \nl
NGC~5033\a   &131327.5 $+$363538& 2.027&  9.88\p 0.57& 23.92\p 0.69&  5.23\p 0.31\nl
NGC~5055\a   &131549.3 $+$420146& 0.528&  7.12\p 1.00& 11.70\p 0.49&  6.07\p 0.12\nl
NGC~5194\a   &132952.7 $+$471143& 0.523& 11.20\p 0.87& 18.40\p 0.57&  2.60\p 0.30\nl
NGC~5195\a   &132959.6 $+$471559& 0.556& 17.30\p 1.30& 41.47\p 0.75&  4.88\p 0.78\nl
Tololo~89\c  &140121.7 $-$330346& 2.403&\nodata      &\nodata      &\nodata      \nl
NGC~5408\c   &140321.0 $-$412244& 0.182&\nodata      &\nodata      &\nodata      \nl
NGC~5474     &140501.5 $+$533945& 0.367&\nodata      &\nodata      &\nodata      \nl
NGC~5713     &144011.5 $-$001720& 6.833&  8.90\p 1.32& 24.29\p 0.76&  4.57\p 0.71\nl
NGC~5866\a   &150629.5 $+$554548& 1.817&  5.51\p 0.87& 13.54\p 0.51&  2.53\p 0.17\nl
 IC~4710     &182838.1 $-$665854& 0.635&\nodata      &\nodata      &\nodata      \nl
NGC~6946\b   &203452.3 $+$600914& 0.907& 18.16\p 1.05& 37.88\p 0.86&  8.77\p 1.81\nl
NGC~7331\a   &223704.1 $+$342456& 1.820&  2.90\p 0.40&  8.76\p 0.51&  2.90\p 0.20\nl
NGC~7552\a\b &231610.8 $-$423505& 8.677& 18.98\p 0.82& 36.53\p 1.51& 14.89\p 5.45\nl
NGC~7793     &235749.8 $-$323527& 0.127&  2.30\p 0.70&  2.33\p 0.43&  0.88\p 0.27\nl
NGC0628~00   &013645.1 $+$154751& 0.464&\nodata      &  2.12\p 0.46&  1.60\p 0.30\nl
NGC0628~01   &013637.5 $+$154512& 0.463&\nodata      &  1.48\p 0.48&  1.02\p 0.36\nl
NGC0628~02   &013638.8 $+$154425& 0.464&\nodata      &  1.82\p 0.60&  0.84\p 0.33\nl
NGC0628~03   &013635.5 $+$155011& 0.463&\nodata      &\nodata      &\nodata      \nl
NGC1097~00   &024624.0 $-$301751& 2.313&  2.19\p 0.84&  3.42\p 0.54&  1.47\p 0.50\nl
NGC1097~01   &024615.2 $-$301511& 2.315&\nodata      &  3.65\p 0.64&  1.82\p 0.31\nl
NGC1097~02   &024614.2 $-$301459& 2.317&\nodata      &  2.36\p 0.50&  1.45\p 0.27\nl
NGC1566~00   &042003.1 $-$545631& 2.313&\nodata      &  4.47\p 0.75&  1.88\p 0.53\nl
NGC1566~01   &041955.5 $-$545614& 2.312&  2.66\p 1.48&  6.48\p 0.67&  2.92\p 0.62\nl
NGC1566~02   &041957.9 $-$545508& 2.313&  1.83\p 0.64&  5.62\p 0.78&  2.55\p 0.34\nl
NGC2403~06   &073645.5 $+$653700& 0.082&\nodata      &\nodata      &\nodata      \nl
NGC2403~07   &073652.7 $+$653646& 0.082&\nodata      &\nodata      &  0.89\p 0.17\nl
NGC2403~08   &073706.9 $+$653639& 0.082&  0.76\p 0.38&  2.05\p 0.20&\nodata      \nl
NGC2403~09   &073717.9 $+$653346& 0.082&\nodata      &\nodata      &  0.56\p 0.20\nl
NGC2403~10   &073619.5 $+$653704& 0.082&\nodata      &\nodata      &\nodata      \nl
NGC2403~11   &073628.5 $+$653350& 0.082&\nodata      &\nodata      &\nodata      \nl
HolmbII~00   &081913.3 $+$704308& 0.111&\nodata      &\nodata      &\nodata      \nl
HolmbII~01   &081927.0 $+$704159& 0.111&\nodata      &\nodata      &\nodata      \nl
HolmbII~02   &081928.8 $+$704221& 0.111&\nodata      &\nodata      &\nodata      \nl
HolmbII~03   &081850.1 $+$704448& 0.111&\nodata      &\nodata      &\nodata      \nl
HolmbII~04   &081928.9 $+$704301& 0.102&\nodata      &\nodata      &\nodata      \nl
NGC2976~00   &094707.8 $+$675552& 0.101&  2.49\p 0.98&  3.53\p 0.70&  2.26\p 0.49\nl
NGC2976~01   &094724.1 $+$675357& 0.101&  2.45\p 1.00&  4.37\p 0.72&  2.50\p 0.32\nl
NGC3031~00   &095600.4 $+$690402& 0.114&\nodata      &  1.19\p 0.44&  0.61\p 0.16\nl
NGC3031~01   &095540.7 $+$685945& 0.105&\nodata      &  1.32\p 0.47&  0.52\p 0.26\nl
NGC3031~02   &095524.4 $+$690815& 0.114&\nodata      &  1.52\p 0.27&  0.53\p 0.18\nl
NGC3031~03   &095553.2 $+$685904& 0.122&\nodata      &  1.22\p 0.33&  0.57\p 0.07\nl
NGC3031~04   &095456.6 $+$690847& 0.133&\nodata      &  0.43\p 0.20&  0.29\p 0.00\nl
NGC3031~05   &095442.3 $+$690336& 0.114&\nodata      &  1.79\p 0.48&  0.67\p 0.26\nl
NGC3031~06   &095617.4 $+$684950& 0.105&\nodata      &\nodata      &\nodata      \nl
IC~2574~00   &102848.4 $+$682802& 0.034&\nodata      &\nodata      &\nodata      \nl
NGC3521~00   &110546.3 $-$000410& 0.801&\nodata      &  1.78\p 0.94&  0.53\p 0.12\nl
NGC3521~01   &110550.0 $-$000340& 0.799&\nodata      &  6.67\p 0.78&  3.93\p 0.31\nl
NGC3521~02   &110549.2 $-$000232& 0.800&\nodata      & 11.00\p 0.90&  4.18\p 0.54\nl
NGC3521~03   &110547.7 $+$000033& 0.801&\nodata      &  2.04\p 0.22&  0.98\p 0.36\nl
NGC3627~00   &112016.3 $+$125750& 0.695&  3.72\p 1.27&  7.40\p 0.77&  2.95\p 0.56\nl
NGC3627~01   &112016.4 $+$125844& 0.697&  9.42\p 0.61& 22.73\p 0.80&  4.03\p 0.51\nl
NGC3627~02   &112016.1 $+$125952& 0.702&\nodata      &  5.63\p 1.04&  2.08\p 0.30\nl
NGC3938~00   &115246.4 $+$440701& 1.414&\nodata      &  3.14\p 0.50&  0.94\p 0.16\nl
NGC3938~01   &115300.2 $+$440749& 1.408&\nodata      &\nodata      &\nodata      \nl
NGC3938~02   &115300.0 $+$440801& 1.410&\nodata      &\nodata      &\nodata      \nl
NGC4254~00   &121849.1 $+$142359&11.017&\nodata      &  3.61\p 1.13&  1.91\p 0.17\nl
NGC4254~01   &121844.7 $+$142425&11.022&\nodata      &  2.20\p 0.84&  1.00\p 0.34\nl
NGC4321~00   &122259.0 $+$154935& 1.609&\nodata      &  3.49\p 0.76&  1.62\p 0.39\nl
NGC4321~01   &122247.4 $+$154945& 1.612&\nodata      &  2.10\p 0.46&  1.52\p 0.27\nl
NGC4321~02   &122249.9 $+$155029& 1.626&\nodata      &  2.17\p 0.66&  1.18\p 0.31\nl
NGC4631~00   &124140.8 $+$323151& 0.514&\nodata      &\nodata      &  0.92\p 0.50\nl
NGC4631~01   &124210.9 $+$323236& 0.518& 12.23\p 2.35& 23.73\p 0.83& 10.09\p 1.05\nl
NGC4631~02   &124221.4 $+$323307& 0.515&\nodata      &\nodata      &  0.89\p 0.22\nl
NGC4736~00   &125049.6 $+$410723& 0.199&  2.89\p 0.62&  7.99\p 0.59&  2.36\p 0.50\nl
NGC4736~01   &125049.6 $+$410734& 0.199&  3.35\p 0.67&  9.02\p 0.34&  2.96\p 0.50\nl
NGC4736~02   &125056.3 $+$410720& 0.199&  5.83\p 0.66& 10.38\p 0.49&  2.45\p 0.73\nl
NGC5055~00   &131558.1 $+$420026& 0.484&\nodata      &  3.07\p 0.60&  2.19\p 0.23\nl
NGC5194~00   &132953.1 $+$471240& 0.482&  3.13\p 0.59&  7.33\p 0.46&  2.71\p 0.47\nl
NGC5194~01   &132944.1 $+$471021& 0.478&  4.70\p 0.81&  8.65\p 0.42&  3.01\p 0.58\nl
NGC5194~02   &132944.6 $+$470955& 0.480&  2.83\p 1.00&  4.76\p 0.43&  1.82\p 0.60\nl
NGC5194~03   &132956.2 $+$471407& 0.478&  2.01\p 0.46&  5.53\p 0.38&  2.09\p 0.34\nl
NGC5194~04   &132959.6 $+$471401& 0.478&  2.39\p 0.52&  4.37\p 0.45&  1.38\p 0.34\nl
NGC5194~05   &132939.5 $+$470835& 0.477&  1.45\p 1.21&  4.76\p 0.47&  2.43\p 0.39\nl
NGC5194~06   &133002.5 $+$470952& 0.478&\nodata      &  3.94\p 0.66&  2.06\p 0.61\nl
NGC5194~07   &133001.6 $+$471252& 0.478&  1.63\p 0.65&  6.98\p 0.52&  2.99\p 1.38\nl
NGC5194~08   &132959.9 $+$471112& 0.476&\nodata      &  6.18\p 0.57&  2.93\p 0.57\nl
NGC5194~09   &132956.7 $+$471046& 0.478&  3.07\p 0.75&  9.78\p 0.39&  3.97\p 0.40\nl
NGC5194~10   &132949.7 $+$471329& 0.478&\nodata      &  2.46\p 0.34&  2.15\p 0.20\nl
Tololo89~00  &140120.2 $-$330410& 2.416&\nodata      &  1.55\p 0.49&\nodata      \nl
NGC5408~00   &140318.3 $-$412252& 0.183&\nodata      &\nodata      &\nodata      \nl
NGC5713~00   &144012.2 $-$001747& 6.833&  1.59\p 0.10&  3.07\p 0.69&  0.92\p 0.22\nl
NGC5713~01   &144010.6 $-$001747& 6.859&  3.77\p 0.69&  5.38\p 1.06&  1.66\p 0.36\nl
NGC6822~00   &194452.9 $-$144311& 0.002&  2.19\p 0.48&  2.81\p 0.36&  1.79\p 0.42\nl
NGC6822~01   &194505.2 $-$144313& 0.002&  2.11\p 0.47&\nodata      &\nodata      \nl
NGC6822~02   &194431.6 $-$144201& 0.002&\nodata      &  1.37\p 0.61&  0.69\p 0.37\nl
NGC6822~03   &194434.1 $-$144222& 0.002&\nodata      &  1.00\p 0.32&\nodata      \nl
NGC6822~04   &194448.6 $-$145227& 0.002&\nodata      &\nodata      &  1.02\p 0.32\nl
NGC6822~05   &194450.6 $-$145251& 0.002&\nodata      &  1.20\p 0.52&  1.81\p 0.31\nl
NGC6822~06   &194457.2 $-$144749& 0.002&\nodata      &\nodata      &  0.98\p 0.37\nl
NGC6946~00   &203516.7 $+$601057& 0.367&\nodata      &  2.64\p 0.25&  1.39\p 0.30\nl
NGC6946~01   &203525.1 $+$601003& 0.467&  1.35\p 0.30&  1.35\p 0.42&  1.10\p 0.36\nl
NGC6946~02   &203452.3 $+$601241& 0.401&  1.47\p 0.46&  2.28\p 0.36&  1.61\p 0.33\nl
NGC6946~03   &203419.5 $+$601009& 0.400&  2.29\p 0.20&  2.53\p 0.47&  1.59\p 0.23\nl
NGC6946~04   &203439.0 $+$600453& 0.400&  1.09\p 0.19&  0.95\p 0.23&  0.77\p 0.11\nl
NGC6946~05   &203505.8 $+$601059& 0.365&  4.65\p 0.58&  8.01\p 0.37&  3.45\p 0.70\nl
NGC6946~06   &203511.0 $+$600859& 0.367&  4.66\p 0.99&  7.73\p 0.60&  3.49\p 0.83\nl
NGC6946~07   &203432.3 $+$601020& 0.366&  3.81\p 2.38&  7.09\p 0.52&  3.41\p 0.47\nl
NGC6946~08   &203513.0 $+$600851& 0.366&  5.17\p 0.68&  7.04\p 0.47&  3.91\p 0.60\nl
NGC7793~00   &235748.8 $-$323659& 0.126&\nodata      &\nodata      &\nodata      \nl
NGC7793~01   &235741.1 $-$323551& 0.116&\nodata      &\nodata      &  0.37\p 0.08\nl
NGC7793~02   &235756.1 $-$323540& 0.127&\nodata      &\nodata      &  0.47\p 0.28\nl
NGC7793~03   &235748.9 $-$323453& 0.126&\nodata      &\nodata      &  0.32\p 0.29\nl

\enddata
\tablecomments{\footnotesize Fluxes and their statistical uncertainties are listed in units of 10$^{-9}$~W~m$^{-2}$~sr$^{-1}$.  Calibration uncertainties are an additional $\sim$25\%.  The 3$\sigma$ upper limits are provided for nondetections in several key lines.  The nuclear targets are listed first followed by the extranuclear targets.}
\tablecomments{\footnotesize The nuclei for eight SINGS galaxies are not listed in this table.  No high resolution infrared spectral data were taken for the optical centers of Holmberg~I, Holmberg~II, M~81~Dwarf~A, M~81~Dwarf~B, IC~2574, DDO~154, NGC~3034 (M~82), and NGC~6822.}
\tablenotetext{a}{\footnotesize Considered to harbor an AGN, based on optical spectroscopy (Moustakas et al. 2009).}
\tablenotetext{b}{\footnotesize The Short-High maps are 45\arcsec$\times$33\arcsec\ instead of the standard $\sim$23\arcsec$\times$15\arcsec.}
\tablenotetext{c}{\footnotesize The infrared emission peaks outside of the field of view of the nuclear spectral maps.  The infrared center was thus observed as an extranuclear target, and the data from those observations are listed further down the table.}
\end{deluxetable}

\begin{deluxetable}{lrrrrrrrrr}
\rotate
\def\a{\tablenotemark{a}}
\def\b{\tablenotemark{b}}
\def\c{\tablenotemark{c}}
\def\p{$\pm$}
\tabletypesize{\scriptsize}
\tablenum{2}
\label{tab:fluxes2}
\tablecaption{Nuclear and ExtraNuclear Emission Line Fluxes: Ionic Forbidden}
\tablewidth{0pc}
\tablehead{
\colhead{Target} &
\colhead{[SIV]} &
\colhead{[NeII]} &
\colhead{[NeV]} &
\colhead{[NeIII]} &
\colhead{[SIII]} &
\colhead{[OIV]} &
\colhead{[FeII]} &
\colhead{[SIII]} &
\colhead{[SiII]} 
\\
\colhead{} &
\colhead{10.51$\mu$m} &
\colhead{12.81$\mu$m} &
\colhead{14.32$\mu$m} &
\colhead{15.56$\mu$m} &
\colhead{18.71$\mu$m} &
\colhead{25.89$\mu$m} &
\colhead{25.99$\mu$m} &
\colhead{33.48$\mu$m} &
\colhead{34.82$\mu$m} 
\\
\colhead{} & 
\colhead{34.8~eV} &
\colhead{21.6~eV} &
\colhead{97.1~eV} &
\colhead{41.0~eV} &
\colhead{23.3~eV} &
\colhead{54.9~eV} &
\colhead{~7.9~eV} &
\colhead{23.3~eV} &
\colhead{~8.2~eV} 
}
\startdata
NGC~0024   &$<$  0.82    &  3.84\p 0.49&$<$  1.09    &  1.64\p 0.50&  1.72\p 0.80&$<$  0.52    &$<$  0.28    &  2.56\p 0.39&  3.94\p 0.29\nl
NGC~0337   &  1.93\p 0.78& 23.13\p 0.50&$<$  0.91    &  9.90\p 0.37& 15.70\p 0.92&$<$  0.60    &  1.13\p 0.24& 24.76\p 0.75& 25.90\p 0.59\nl
NGC~0584   &$<$  1.15    &$<$  1.09    &$<$  1.14    &  2.10\p 0.64&  1.20\p 0.45&$<$  0.37    &$<$  0.57    &$<$  0.90    &  1.62\p 0.43\nl
NGC~0628   &$<$  2.17    &  7.69\p 0.43&$<$  1.23    &$<$  1.17    &  3.11\p 0.59&$<$  0.47    &$<$  0.21    &  4.98\p 0.42&  7.58\p 0.58\nl
NGC~0855   &  2.51\p 0.31&  6.82\p 0.51&$<$  0.89    &$<$  1.23    &  9.26\p 0.54&$<$  0.49    &$<$  0.81    & 10.67\p 0.52&  9.93\p 0.52\nl
NGC~0925   &$<$  2.03    & 12.80\p 0.70&$<$  1.25    &  6.29\p 0.58&  5.80\p 0.87&$<$  0.45    &  0.68\p 0.21& 11.19\p 0.37& 12.46\p 0.49\nl
NGC~1097\b &$<$  3.66    &295.78\p 5.89&$<$  1.00    & 25.08\p 0.92& 77.09\p 0.92&  2.61\p 0.46& 13.77\p 1.40&119.01\p 2.65&280.39\p 3.05\nl
NGC~1266   &$<$  2.22    & 35.74\p 1.20&$<$  1.13    & 11.90\p 0.55&  1.47\p 0.54&$<$  1.75    &  3.75\p 1.14&  4.99\p 0.58& 19.87\p 3.58\nl
NGC~1291   &$<$  2.08    &  6.45\p 0.72&$<$  1.06    &  8.89\p 0.50&  3.35\p 0.68&  1.75\p 0.31&  1.55\p 0.25&  3.84\p 0.67& 10.70\p 0.51\nl
NGC~1316   &$<$  2.09    & 15.70\p 1.18&$<$  0.75    & 13.43\p 0.47&  3.14\p 0.60&  3.16\p 0.18&  3.91\p 0.27&  3.91\p 0.83& 14.87\p 0.78\nl
NGC~1377   &$<$  2.57    &  5.01\p 1.29&$<$  2.32    &  3.18\p 0.68&  1.28\p 0.54&$<$  2.12    &$<$  2.82    &$<$  3.38    &$<$  5.16    \nl
NGC~1404   &$<$  1.75    &  1.67\p 0.45&$<$  1.19    &  1.23\p 0.34&$<$  1.09    &$<$  0.43    &$<$  0.43    &$<$  0.54    &$<$  1.04    \nl
NGC~1482\b &$<$  5.16    &231.78\p 4.32&$<$  1.33    & 28.48\p 0.32& 61.16\p 0.77&$<$  3.25    &  8.77\p 1.33&117.12\p 3.92&214.68\p 3.11\nl
NGC~1512\b &$<$  1.24    & 15.69\p 3.51&$<$  0.55    &  2.38\p 0.40&  6.61\p 0.63&$<$  0.42    &  0.79\p 0.23& 11.53\p 0.57& 21.69\p 0.49\nl
NGC~1566   &$<$  2.50    & 21.09\p 1.04&  1.19\p 0.35& 11.53\p 0.44&  8.55\p 0.54&  7.50\p 0.95&  1.56\p 0.31&  9.03\p 0.59& 18.37\p 0.87\nl
NGC~1705\b &  3.49\p 0.35&  1.64\p 0.39&$<$  0.46    &  6.31\p 0.56&  3.73\p 0.51&  1.25\p 0.19&$<$  0.44    &  6.10\p 0.64&  4.76\p 0.51\nl
NGC~2403   &$<$  1.89    &  6.29\p 0.74&$<$  0.88    &  2.59\p 0.33&  4.26\p 0.49&$<$  0.44    &  0.57\p 0.19&  7.33\p 0.39& 11.94\p 1.32\nl
DDO~053    &$<$  1.44    &$<$  1.40    &$<$  1.10    &  1.21\p 0.15&$<$  1.11    &$<$  0.46    &$<$  0.43    &$<$  1.19    &$<$  1.33    \nl
NGC~2798   &  6.79\p 1.21&265.44\p 5.53&$<$  2.37    & 41.26\p 0.52& 99.05\p 1.16&$<$  6.28    &  8.86\p 2.76& 98.69\p 2.94&128.81\p 2.93\nl
NGC~2841   &$<$  1.64    &  7.03\p 0.45&$<$  5.51    &  8.30\p 0.61&  3.42\p 0.63&  0.96\p 0.17&  0.96\p 0.18&  4.88\p 0.29& 10.98\p 0.44\nl
NGC~2915   &  2.33\p 0.22&  3.93\p 0.41&$<$  0.92    & 13.40\p 0.52&  5.38\p 0.51&  0.55\p 0.16&  0.25\p 0.07&  7.75\p 0.22&  6.06\p 0.51\nl
NGC~2976   &$<$  1.94    &  9.17\p 0.55&$<$  1.22    &  3.41\p 0.33&  7.56\p 0.48&$<$  0.38    &  0.52\p 0.15& 10.84\p 0.41& 10.75\p 0.47\nl
NGC~3049   &$<$  1.80    & 47.94\p 1.34&$<$  1.10    &  7.35\p 0.40& 31.72\p 0.49&$<$  1.23    &  0.88\p 0.24& 32.27\p 0.70& 26.29\p 0.79\nl
NGC~3031   &  2.03\p 1.53& 38.43\p 0.98&$<$  0.93    & 30.34\p 0.70&  9.65\p 0.73&  5.72\p 1.10&  4.22\p 0.46&  9.60\p 0.65& 29.67\p 0.72\nl
Holmberg~IX&$<$  1.62    &$<$  1.10    &$<$  0.79    &$<$  0.93    &$<$  1.01    &$<$  0.47    &$<$  0.51    &$<$  1.03    &$<$  1.25    \nl
NGC~3190   &$<$  2.58    & 10.18\p 0.86&$<$  0.82    &  5.93\p 0.42&  3.09\p 0.66&  1.15\p 0.28&  1.63\p 0.22&  2.62\p 0.42&  8.95\p 0.42\nl
NGC~3184   &$<$  1.71    & 22.31\p 1.13&$<$  1.02    &  2.66\p 0.51& 10.11\p 0.57&$<$  0.36    &  1.06\p 0.27&  9.82\p 0.41& 14.93\p 0.31\nl
NGC~3198   &$<$  2.05    & 18.24\p 0.64&$<$  1.23    &  1.00\p 0.41&  6.52\p 1.20&$<$  0.39    &  1.05\p 0.21&  7.80\p 0.68& 13.41\p 0.62\nl
NGC~3265   &$<$  2.47    & 36.68\p 0.91&$<$  0.96    &  6.72\p 0.55& 18.99\p 0.75&$<$  0.88    &  1.36\p 0.49& 17.83\p 0.71& 20.92\p 0.81\nl
Markarian33& 19.37\p 0.94& 73.32\p 0.74&$<$  1.27    & 55.21\p 0.48& 62.70\p 0.90&  1.10\p 0.34&  2.50\p 0.51& 54.33\p 1.07& 34.99\p 1.01\nl
NGC~3351\b &$<$  1.96    &110.94\p 5.29&$<$  1.16    &  9.37\p 0.29& 39.66\p 0.67&  1.29\p 0.62&  4.90\p 0.71& 54.77\p 1.74&105.81\p 1.40\nl
NGC~3521   &$<$  2.01    & 16.92\p 0.66&$<$  1.17    &  9.08\p 0.47&  2.88\p 0.80&  2.76\p 1.15&  1.25\p 0.21&  9.06\p 0.81& 23.29\p 0.65\nl
NGC~3621   &  2.01\p 0.49& 19.57\p 0.86&  1.28\p 0.87&  6.98\p 0.40&  8.60\p 0.77&  5.66\p 0.13&  1.43\p 0.32& 14.84\p 0.43& 23.02\p 0.42\nl
NGC~3627   &$<$  3.25    & 28.60\p 1.28&$<$  1.14    & 10.48\p 0.60&  6.12\p 0.57&  2.04\p 0.94&  3.81\p 0.20&  9.76\p 0.64& 24.85\p 1.14\nl
NGC~3773   &  4.61\p 0.75& 21.74\p 0.69&$<$  1.00    & 19.19\p 0.43& 18.16\p 0.66&  0.44\p 0.83&  1.14\p 0.26& 22.89\p 0.55& 16.17\p 0.50\nl
NGC~3938   &$<$  1.40    &  6.69\p 0.60&$<$  0.81    &  1.25\p 0.17&  1.53\p 0.39&  0.31\p 0.18&  0.47\p 0.09&  3.53\p 0.49&  9.29\p 0.49\nl
NGC~4125   &$<$  1.57    &  2.89\p 0.59&$<$  0.76    &  4.06\p 0.58&  1.14\p 0.60&  0.93\p 0.26&  1.47\p 0.40&  1.34\p 0.36&  5.92\p 0.45\nl
NGC~4236   &$<$  1.64    &$<$  1.18    &$<$  1.87    &$<$  0.91    &$<$  1.00    &$<$  0.28    &$<$  0.96    &$<$  0.78    &$<$  1.06    \nl
NGC~4254   &$<$  2.49    & 65.51\p 1.18&$<$  1.13    &  7.51\p 0.47& 12.17\p 0.66&  3.11\p 0.23&  2.74\p 0.19& 22.01\p 0.56& 61.07\p 0.57\nl
NGC~4321\b &$<$  1.94    & 76.96\p 1.45&$<$  0.63    &  9.22\p 0.30& 16.17\p 0.54&$<$  0.94    &  4.39\p 1.27& 27.55\p 0.81& 79.50\p 0.98\nl
NGC~4450   &$<$  1.66    &  4.31\p 0.82&$<$  0.65    &  2.60\p 0.46&  1.42\p 0.39&  0.78\p 0.22&  0.85\p 0.37&  1.93\p 0.85&  4.68\p 0.63\nl
NGC~4536\b &$<$  1.94    &195.88\p 1.58&$<$  0.63    & 30.41\p 0.57& 80.37\p 0.86&$<$  4.02    &  9.04\p 2.24&141.25\p 2.03&174.73\p 2.59\nl
NGC~4552   &$<$  1.90    &  2.06\p 0.95&$<$  1.10    &  3.63\p 0.61&$<$  1.52    &$<$  0.65    &$<$  0.48    &  1.60\p 0.51&  1.64\p 0.43\nl
NGC~4559   &$<$  1.94    & 10.25\p 0.66&$<$  1.26    &  2.83\p 0.32&  5.73\p 0.55&  0.15\p 0.08&  0.68\p 0.12& 11.55\p 0.32& 12.62\p 0.49\nl
NGC~4569   &$<$  2.26    & 44.51\p 1.16&$<$  1.15    & 19.68\p 0.50&  9.49\p 1.11&  3.52\p 0.42&  4.28\p 0.25&  9.35\p 0.98& 36.65\p 1.37\nl
NGC~4579   &$<$  1.56    & 29.99\p 0.92&$<$  0.90    & 14.41\p 0.43&  5.58\p 1.11&  3.38\p 0.11&  5.44\p 0.41&  4.87\p 0.45& 23.48\p 0.56\nl
NGC~4594   &  3.64\p 0.80& 17.29\p 0.97&$<$  1.89    & 20.05\p 0.71&  6.15\p 0.71&  2.75\p 0.29&  3.14\p 0.28&  8.67\p 0.53& 17.06\p 0.55\nl
NGC~4625   &$<$  2.08    &  7.31\p 0.60&$<$  1.03    &  1.49\p 0.45&  3.58\p 0.53&$<$  0.32    &  0.45\p 0.17&  5.49\p 0.48&  8.35\p 0.69\nl
NGC~4631   &  3.90\p 0.56&164.45\p 2.21&$<$  0.64    & 36.55\p 0.54& 81.06\p 0.97&  3.64\p 1.23&  7.31\p 1.06&149.01\p 0.97&192.18\p 2.00\nl
NGC~4725   &  1.38\p 0.59&  2.79\p 0.55&$<$  0.98    &  3.27\p 0.38&$<$  0.93    &  1.90\p 0.32&  1.62\p 0.20&  1.77\p 0.36&  3.99\p 0.45\nl
NGC~4736   &$<$  5.25    & 16.63\p 0.81&$<$  1.00    & 17.28\p 0.58&  8.02\p 1.07&  4.08\p 0.50&  6.52\p 0.48&  8.89\p 1.38& 28.87\p 2.86\nl
NGC~4826   &$<$  4.09    &129.26\p 2.57&$<$  1.59    & 28.13\p 0.52& 50.62\p 0.49&  5.26\p 0.06& 11.76\p 0.71& 71.46\p 1.27&130.41\p 1.31\nl
 DDO~165   &$<$  1.43    &$<$  0.86    &$<$  0.93    &$<$  0.78    &$<$  1.16    &$<$  0.30    &$<$  0.55    &$<$  0.55    &$<$  0.99    \nl
NGC~5033   &  2.83\p 0.31& 64.67\p 1.80&  1.40\p 0.30& 19.59\p 0.52& 14.88\p 0.50&  9.98\p 0.74&  5.44\p 0.46& 27.56\p 0.41& 72.19\p 0.60\nl
NGC~5055   &$<$  2.30    & 27.32\p 0.95&$<$  1.17    & 11.80\p 0.35& 21.06\p 0.51&  2.86\p 0.11&  3.38\p 0.24& 15.67\p 0.38& 45.91\p 0.55\nl
NGC~5194   &  5.70\p 0.95& 83.20\p 1.23&  2.68\p 0.19& 43.93\p 0.64& 15.35\p 0.76& 19.79\p 0.72&  8.48\p 1.22& 22.51\p 0.53& 73.72\p 0.77\nl
NGC~5195   &$<$  2.90    & 21.88\p 0.80&$<$  1.24    &  8.55\p 0.47&  3.28\p 0.80&$<$  2.36    &  1.82\p 0.73&  4.13\p 1.06& 16.20\p 1.12\nl
Tololo~89\a&$<$  1.85    &$<$  1.06    &$<$  0.62    &$<$  1.30    &$<$  1.24    &$<$  0.38    &$<$  0.21    &  1.48\p 0.50&$<$  1.43    \nl
NGC~5408\a &$<$  1.39    &$<$  1.10    &$<$  1.15    &  1.49\p 0.47&  1.54\p 0.30&$<$  0.63    &$<$  0.63    &  2.56\p 0.50&  2.36\p 0.46\nl
NGC~5474   &$<$  1.34    &$<$  1.00    &$<$  1.02    &$<$  1.31    &$<$  0.61    &$<$  0.38    &$<$  0.37    &  1.88\p 0.25&  3.12\p 0.31\nl
NGC~5713   &  1.56\p 0.86&156.77\p 2.35&$<$  0.99    & 20.91\p 0.39& 57.53\p 0.75&  3.43\p 0.77&  7.80\p 0.85& 81.00\p 0.98&111.06\p 1.24\nl
NGC~5866   &$<$  1.86    &  9.56\p 1.03&$<$  1.19    &  6.03\p 0.47&  1.52\p 0.70&  1.15\p 0.21&  1.75\p 0.14&  5.41\p 0.33& 12.38\p 0.29\nl
 IC~4710   &  4.30\p 0.92&  1.08\p 0.92&$<$  1.28    &  5.33\p 0.41&  4.11\p 0.40&$<$  0.37    &$<$  0.38    &  5.96\p 0.46&  1.67\p 0.26\nl
NGC~6946\b &$<$  2.86    &218.43\p 4.74&$<$  0.77    & 19.91\p 0.53& 63.54\p 0.92&$<$  3.65    &  9.71\p 2.05& 85.72\p 3.05&161.71\p 2.21\nl
NGC~7331   &$<$  1.75    & 23.71\p 0.93&$<$  0.64    & 12.52\p 0.44&  7.14\p 0.61&  3.69\p 0.24&  2.63\p 0.21& 16.64\p 0.33& 40.27\p 0.51\nl
NGC~7552\b &$<$  1.70    &423.29\p19.65&$<$  1.82    & 35.54\p 0.51&129.25\p 1.23&$<$  5.27    & 16.41\p 3.44&140.40\p 5.56&207.63\p 5.97\nl
NGC~7793   &$<$  2.38    & 12.69\p 0.73&$<$  1.24    &  3.25\p 0.46&  9.92\p 0.67&$<$  0.40    &  0.71\p 0.21& 11.52\p 0.28& 11.62\p 0.45\nl
NGC0628~00 &$<$  1.86    & 32.69\p 0.71&$<$  1.52    &  3.04\p 0.53& 24.73\p 0.73&$<$  1.00    &$<$  1.17    & 27.42\p 0.50& 14.18\p 0.44\nl
NGC0628~01 &$<$  2.11    & 15.77\p 0.45&$<$  1.45    &  2.83\p 0.38&  9.82\p 0.59&$<$  0.70    &$<$  0.68    & 10.37\p 0.45&  7.79\p 0.70\nl
NGC0628~02 &  5.96\p 0.49& 14.03\p 0.52&$<$  1.04    & 11.53\p 0.41& 12.85\p 0.58&$<$  0.78    &  0.46\p 0.13& 16.46\p 1.24&  9.85\p 0.83\nl
NGC0628~03 &$<$  2.12    &  5.65\p 0.56&$<$  1.18    &  2.44\p 0.48&  6.38\p 0.55&$<$  0.62    &$<$  0.84    &  5.99\p 0.48&  3.97\p 0.51\nl
NGC1097~00 &$<$  1.90    & 13.90\p 0.65&$<$  1.19    &  1.95\p 0.52&  5.95\p 1.16&$<$  0.72    &  0.55\p 0.15&  6.65\p 1.10& 12.96\p 0.88\nl
NGC1097~01 &$<$  1.98    &  5.81\p 0.54&$<$  1.51    &$<$  1.36    &$<$  1.59    &$<$  0.72    &  0.48\p 0.20&  5.65\p 0.87&  7.18\p 0.81\nl
NGC1097~02 &$<$  2.25    &  7.69\p 0.76&$<$  1.32    &  2.68\p 0.31&  3.58\p 0.33&$<$  0.59    &$<$  0.59    &  7.31\p 0.73&  8.47\p 1.01\nl
NGC1566~00 &$<$  2.81    & 22.27\p 1.13&$<$  1.72    &  3.92\p 0.46&  8.30\p 0.75&$<$  0.75    &  0.61\p 0.14& 13.77\p 0.73& 23.80\p 0.98\nl
NGC1566~01 &  0.66\p 0.30& 53.18\p 0.78&$<$  1.54    &  8.44\p 0.67& 30.20\p 1.05&$<$  1.38    &$<$  1.36    & 34.66\p 0.87& 38.44\p 1.28\nl
NGC1566~02 &  1.82\p 0.90& 69.94\p 1.14&$<$  1.57    & 11.08\p 0.61& 46.52\p 1.28&$<$  1.41    &  1.12\p 0.35& 49.23\p 1.77& 44.59\p 1.36\nl
NGC2403~06 &  8.86\p 1.40& 57.33\p 1.64&$<$  1.26    & 31.56\p 0.52& 59.57\p 0.73&$<$  0.78    &  1.16\p 0.15& 65.43\p 0.82& 31.19\p 0.82\nl
NGC2403~07 & 10.01\p 1.13& 49.17\p 0.95&$<$  0.99    & 26.64\p 0.70& 51.08\p 0.65&$<$  0.90    &  1.26\p 0.08& 54.68\p 0.84& 27.56\p 0.72\nl
NGC2403~08 & 31.21\p 1.19& 99.96\p 1.78&$<$  1.05    & 88.60\p 0.78&116.16\p 1.71&$<$  0.94    &  2.82\p 0.32&125.66\p 1.09& 59.95\p 1.47\nl
NGC2403~09 &  6.24\p 1.18& 19.46\p 0.86&$<$  1.07    & 18.41\p 0.84& 21.58\p 0.89&$<$  0.55    &  0.43\p 0.18& 25.92\p 0.66& 11.83\p 0.57\nl
NGC2403~10 &  8.54\p 1.01& 42.95\p 0.97&$<$  1.28    & 23.21\p 0.70& 42.55\p 1.11&$<$  0.80    &  1.16\p 0.15& 47.66\p 0.56& 22.71\p 0.61\nl
NGC2403~11 &  3.50\p 0.43&  8.68\p 0.41&$<$  1.24    & 12.83\p 0.78& 10.60\p 0.76&$<$  0.63    &$<$  0.38    & 11.28\p 0.59&  4.78\p 0.96\nl
HolmbII~00 &  1.29\p 0.27&$<$  1.03    &$<$  0.85    &  4.10\p 0.65&  4.35\p 0.88&$<$  0.41    &  0.45\p 0.13&  3.64\p 0.84&  3.06\p 0.48\nl
HolmbII~01 &$<$  1.93    &$<$  1.15    &$<$  0.87    &  1.39\p 0.37&$<$  1.17    &$<$  0.38    &$<$  0.47    &$<$  0.72    &$<$  1.09    \nl
HolmbII~02 &$<$  2.18    &$<$  0.98    &$<$  0.82    &  1.67\p 0.52&  1.13\p 0.60&  1.25\p 0.80&$<$  0.35    &  0.90\p 0.23&  1.57\p 0.47\nl
HolmbII~03 &  1.02\p 0.60&$<$  1.05    &$<$  1.16    &  1.49\p 0.62&  1.02\p 0.30&$<$  0.43    &$<$  0.40    &  0.66\p 0.31&  1.10\p 0.24\nl
HolmbII~04 &$<$  1.81    &$<$  1.31    &$<$  0.88    &  2.49\p 0.58&  0.54\p 0.29&$<$  0.50    &$<$  0.41    &  1.37\p 0.48&  1.81\p 0.49\nl
NGC2976~00 &  8.89\p 1.51& 80.80\p 1.79&$<$  1.50    & 29.58\p 0.68& 90.05\p 1.15&$<$  1.06    &  1.27\p 0.26&100.57\p 1.30& 45.99\p 1.06\nl
NGC2976~01 &  6.84\p 1.64& 43.10\p 1.21&$<$  1.03    & 20.28\p 1.01& 37.82\p 2.17&$<$  0.55    &  0.77\p 0.17& 42.99\p 0.83& 30.52\p 0.93\nl
NGC3031~00 &$<$  1.84    &  5.15\p 0.60&$<$  0.81    &  1.02\p 0.48&  3.12\p 0.56&$<$  0.63    &$<$  0.49    &  4.23\p 0.52&  5.04\p 0.64\nl
NGC3031~01 &$<$  1.92    &  9.58\p 0.60&$<$  1.26    &  4.54\p 0.51&  8.22\p 0.66&$<$  0.42    &$<$  0.74    & 13.05\p 0.55&  7.89\p 0.72\nl
NGC3031~02 &$<$  1.80    &  9.60\p 0.62&$<$  1.05    &  4.57\p 0.56&  8.19\p 0.67&$<$  0.56    &  0.51\p 0.18& 14.88\p 0.46& 11.26\p 0.53\nl
NGC3031~03 &$<$  1.69    & 19.29\p 0.66&$<$  0.93    &  7.58\p 0.52& 18.96\p 0.76&$<$  0.67    &$<$  0.60    & 22.84\p 0.58& 15.44\p 0.63\nl
NGC3031~04 &$<$  1.90    & 13.61\p 0.70&$<$  0.92    &  3.12\p 0.72& 11.36\p 0.71&$<$  0.62    &  0.67\p 0.31& 13.98\p 0.55& 14.09\p 0.60\nl
NGC3031~05 &$<$  1.83    & 11.94\p 0.56&$<$  1.07    &  1.94\p 0.57&  8.67\p 0.70&$<$  0.54    &$<$  0.43    & 10.05\p 0.46&  7.71\p 0.59\nl
NGC3031~06 &$<$  1.40    &$<$  1.28    &$<$  0.78    &  1.86\p 0.42&$<$  1.35    &$<$  0.46    &$<$  0.39    &$<$  0.78    &$<$  0.82    \nl
IC~2574~00 &  4.16\p 0.67&  2.21\p 0.56&$<$  1.67    &  6.95\p 0.15&  3.64\p 0.92&$<$  0.63    &$<$  0.81    &  4.55\p 0.72&  2.14\p 0.63\nl
NGC3521~00 &  4.10\p 1.73&  2.45\p 1.14&$<$  1.57    &  4.76\p 0.53&  3.96\p 1.05&$<$  0.77    &$<$  1.21    &  3.10\p 0.57&  2.64\p 0.67\nl
NGC3521~01 &$<$  3.03    & 23.80\p 1.14&$<$  1.58    &  4.12\p 0.88&  6.36\p 0.82&$<$  1.25    &  1.65\p 0.34& 12.31\p 0.84& 22.46\p 1.07\nl
NGC3521~02 &$<$  3.75    & 36.80\p 1.52&$<$  2.08    &  5.62\p 0.79&  7.87\p 0.72&$<$  0.62    &  1.47\p 0.26& 16.16\p 1.14& 38.58\p 0.99\nl
NGC3521~03 &$<$  2.41    &  5.71\p 0.77&$<$  1.21    &  3.32\p 0.80&  3.33\p 0.30&$<$  0.98    &$<$  0.43    &  4.55\p 0.65&  2.92\p 0.46\nl
NGC3627~00 &$<$  3.03    & 90.75\p 1.54&$<$  2.11    &  8.87\p 0.97& 40.73\p 1.25&$<$  2.02    &  1.46\p 0.49& 49.50\p 1.40& 56.31\p 0.89\nl
NGC3627~01 &$<$  3.64    &249.90\p 4.84&$<$  1.77    & 15.38\p 0.86& 89.02\p 1.14&$<$  2.88    &  6.06\p 0.38&133.24\p 2.00&183.37\p 2.74\nl
NGC3627~02 &$<$  1.96    & 18.44\p 0.94&$<$  1.23    &  2.76\p 0.52&  4.69\p 1.01&  0.61\p 0.22&  0.85\p 0.16&  8.88\p 0.54& 22.46\p 0.81\nl
NGC3938~00 &$<$  2.50    &  6.34\p 1.29&$<$  1.48    &$<$  1.22    &  3.32\p 0.68&$<$  0.62    &  0.42\p 0.07&  3.83\p 0.58&  6.34\p 0.44\nl
NGC3938~01 &  3.59\p 0.83&  3.68\p 0.88&$<$  1.67    &  7.37\p 0.36&  5.10\p 1.25&$<$  0.60    &$<$  0.47    &  4.90\p 0.67&  4.04\p 0.62\nl
NGC3938~02 &  1.93\p 0.66&  2.68\p 0.76&$<$  1.31    &  4.31\p 0.41&  4.82\p 0.72&$<$  0.48    &$<$  0.44    &  4.84\p 0.55&  3.25\p 0.51\nl
NGC4254~00 &$<$  2.73    & 21.24\p 1.07&$<$  1.35    &  4.81\p 0.84&  8.94\p 1.46&$<$  0.76    &$<$  0.92    &  8.89\p 0.69& 16.10\p 0.65\nl
NGC4254~01 &$<$  2.42    &  7.52\p 0.79&$<$  1.39    &  1.82\p 0.58&  4.29\p 1.08&$<$  0.71    &$<$  0.65    &  3.60\p 0.60&  6.70\p 0.88\nl
NGC4321~00 &$<$  2.39    &  7.79\p 0.64&$<$  1.36    &  1.40\p 0.48&  2.29\p 0.79&$<$  0.60    &$<$  0.73    &  4.49\p 0.58&  8.61\p 0.80\nl
NGC4321~01 &$<$  2.22    & 14.29\p 1.00&$<$  1.54    &  3.61\p 0.74&  8.74\p 0.85&$<$  0.68    &$<$  0.89    & 10.74\p 0.81& 11.14\p 0.69\nl
NGC4321~02 &$<$  2.42    & 11.09\p 0.85&$<$  1.34    &  2.78\p 0.54&  4.49\p 0.94&$<$  0.85    &  0.58\p 0.14&  5.27\p 0.97&  9.79\p 0.88\nl
NGC4631~00 &  2.60\p 1.17&  8.11\p 0.75&$<$  1.48    & 10.88\p 1.52&  6.29\p 1.16&  0.96\p 0.36&  0.86\p 0.22&  8.45\p 0.65& 10.28\p 0.68\nl
NGC4631~01 & 14.11\p 1.64&213.84\p 3.97&$<$  2.21    & 68.95\p 1.14&118.14\p 1.39&  3.49\p 1.13&  7.97\p 0.88&173.60\p 1.85&199.30\p 2.52\nl
NGC4631~02 &  2.13\p 1.09& 11.75\p 1.40&$<$  1.56    &  9.38\p 1.10&  9.97\p 1.04&  0.41\p 0.11&  0.61\p 0.09& 13.05\p 0.76& 12.75\p 0.45\nl
NGC4736~00 &  2.30\p 0.42& 65.99\p 1.28&$<$  1.22    & 11.48\p 0.47& 41.98\p 0.78&$<$  1.11    &  2.75\p 0.55& 55.31\p 0.89& 57.72\p 0.69\nl
NGC4736~01 &$<$  2.35    & 62.34\p 1.73&$<$  1.04    & 12.87\p 0.43& 40.26\p 0.64&$<$  1.04    &  2.57\p 0.46& 46.43\p 0.80& 50.36\p 0.53\nl
NGC4736~02 &  8.26\p 0.89& 62.71\p 2.28&$<$  1.17    & 17.12\p 0.57& 51.14\p 0.79&$<$  1.18    &  3.52\p 0.63& 61.74\p 0.66& 60.38\p 0.73\nl
NGC5055~00 &$<$  2.76    & 20.26\p 1.31&$<$  1.48    &  4.26\p 0.63& 10.24\p 1.07&$<$  0.47    &  0.64\p 0.04& 13.58\p 0.55& 15.38\p 1.04\nl
NGC5194~00 &$<$  2.44    & 40.28\p 0.68&$<$  1.00    &  2.70\p 0.50& 12.91\p 0.69&  0.50\p 0.11&  1.21\p 0.30& 22.42\p 0.84& 41.77\p 0.50\nl
NGC5194~01 &$<$  2.94    & 92.62\p 1.11&$<$  1.11    &  4.55\p 0.46& 35.72\p 0.69&$<$  1.13    &  1.58\p 0.42& 46.62\p 0.81& 31.48\p 1.52\nl
NGC5194~02 &$<$  2.36    & 57.60\p 0.86&$<$  1.36    &  6.83\p 0.66& 20.12\p 0.91&$<$  1.11    &  1.44\p 0.43& 32.82\p 0.67& 31.99\p 0.54\nl
NGC5194~03 &$<$  1.67    & 22.50\p 0.64&$<$  1.11    &  3.59\p 0.54&  9.63\p 0.49&  0.99\p 0.29&  1.04\p 0.14& 16.73\p 0.50& 26.72\p 0.50\nl
NGC5194~04 &$<$  2.23    & 49.83\p 1.09&$<$  0.88    &  7.38\p 0.58& 23.79\p 0.51&  0.60\p 0.28&  1.44\p 0.43& 36.57\p 0.43& 46.06\p 0.56\nl
NGC5194~05 &$<$  1.97    & 27.19\p 0.81&$<$  0.75    &  7.03\p 0.38& 13.06\p 0.79&$<$  0.62    &$<$  0.55    & 17.33\p 0.40& 16.12\p 0.46\nl
NGC5194~06 &$<$  3.14    & 46.17\p 1.41&$<$  1.46    &  7.22\p 0.44& 21.83\p 0.67&$<$  0.68    &$<$  0.63    & 27.68\p 0.87& 29.53\p 0.74\nl
NGC5194~07 &$<$  1.94    & 60.50\p 1.33&$<$  1.70    &  5.34\p 0.52& 22.41\p 1.00&  1.27\p 0.09&  2.66\p 0.80& 34.24\p 0.78& 47.43\p 0.93\nl
NGC5194~08 &$<$  2.33    & 37.99\p 1.31&$<$  1.75    &  3.02\p 0.33& 12.53\p 0.76&$<$  0.79    &  0.75\p 0.17& 18.78\p 0.67& 27.84\p 0.78\nl
NGC5194~09 &$<$  1.87    & 34.28\p 0.93&$<$  1.07    &  3.57\p 0.31&  9.37\p 0.85&  0.76\p 0.10&  1.91\p 0.28& 16.92\p 0.58& 35.21\p 0.64\nl
NGC5194~10 &$<$  2.18    &  9.97\p 0.85&$<$  1.59    &  2.57\p 0.73&  2.91\p 0.91&$<$  0.57    &$<$  0.70    &  6.18\p 0.68& 10.59\p 0.76\nl
Tololo89~00& 21.03\p 1.09& 10.55\p 1.05&$<$  1.58    & 34.66\p 0.71& 20.20\p 0.89&  2.37\p 0.40&$<$  0.82    & 24.82\p 0.86&  9.23\p 1.25\nl
NGC5408~00 & 46.51\p 1.09&  6.10\p 1.09&$<$  1.82    & 46.36\p 0.91& 18.14\p 1.10&  6.75\p 0.79&$<$  0.98    & 19.12\p 1.04& 14.34\p 0.82\nl
NGC5713~00 &$<$  2.19    & 13.59\p 0.74&$<$  1.52    &  5.12\p 0.53&  6.40\p 1.46&$<$  1.07    &  0.97\p 0.20& 10.48\p 0.57& 17.05\p 0.67\nl
NGC5713~01 &$<$  2.51    & 35.85\p 0.91&$<$  1.30    &  8.28\p 0.74& 18.72\p 0.94&$<$  1.39    &  1.81\p 0.18& 25.40\p 0.85& 30.86\p 0.82\nl
NGC6822~00 &120.59\p 1.25& 26.89\p 0.71&$<$  1.22    &139.85\p 0.63& 96.95\p 0.87&$<$  2.04    &$<$  1.20    & 90.99\p 0.84& 23.58\p 0.85\nl
NGC6822~01 & 27.98\p 1.10& 14.26\p 0.52&$<$  1.01    & 50.07\p 0.71& 42.20\p 0.67&$<$  0.60    &$<$  0.92    & 43.68\p 0.59&  9.87\p 0.48\nl
NGC6822~02 &  6.24\p 0.86&  4.41\p 0.46&$<$  0.79    & 13.60\p 0.45& 10.61\p 0.58&$<$  0.65    &$<$  0.54    & 14.55\p 0.49&  4.35\p 0.48\nl
NGC6822~03 &$<$  1.71    &  3.62\p 0.48&$<$  1.17    &  4.93\p 0.56&  4.27\p 0.60&$<$  0.56    &$<$  0.65    &  7.84\p 0.41&  6.56\p 0.61\nl
NGC6822~04 &  9.52\p 1.28&  4.56\p 0.71&$<$  1.08    & 13.71\p 0.75&  9.53\p 0.65&$<$  0.50    &$<$  0.49    & 11.15\p 0.38&  9.97\p 0.47\nl
NGC6822~05 &$<$  1.88    &  4.80\p 0.66&$<$  1.29    &  4.19\p 0.39&  4.94\p 0.62&$<$  0.73    &$<$  0.68    &  5.59\p 0.53&  6.40\p 0.79\nl
NGC6822~06 &$<$  2.16    &  2.01\p 0.79&$<$  1.25    &  2.79\p 0.63&  1.85\p 0.74&$<$  0.42    &$<$  0.40    &  2.60\p 0.40&  3.54\p 0.68\nl
NGC6946~00 &  5.04\p 0.90& 21.15\p 0.93&$<$  1.02    & 22.33\p 0.64& 17.51\p 0.47&$<$  0.52    &  0.62\p 0.26& 18.39\p 0.51& 17.30\p 0.55\nl
NGC6946~01 & 11.24\p 1.07& 36.80\p 0.49&$<$  1.19    & 40.78\p 0.62& 37.58\p 0.75&  0.96\p 0.17&  0.88\p 0.25& 49.47\p 0.58& 30.08\p 0.54\nl
NGC6946~02 &  1.91\p 0.34& 23.24\p 0.53&$<$  1.11    & 10.51\p 0.44& 16.46\p 0.62&$<$  0.43    &  0.63\p 0.09& 21.30\p 0.55& 17.38\p 0.39\nl
NGC6946~03 &  1.76\p 0.66& 24.72\p 0.62&$<$  1.11    & 10.41\p 0.46& 17.91\p 0.63&$<$  0.55    &  0.47\p 0.19& 21.82\p 0.43& 18.73\p 0.32\nl
NGC6946~04 &$<$  2.05    & 11.74\p 0.57&$<$  0.85    &  3.59\p 0.43&  7.87\p 0.64&$<$  0.46    &$<$  0.54    & 11.24\p 0.40& 10.01\p 0.41\nl
NGC6946~05 &$<$  2.35    &107.02\p 0.87&$<$  1.59    & 17.82\p 0.46& 56.75\p 0.74&$<$  1.11    &  3.37\p 0.83& 71.92\p 0.95& 77.77\p 1.17\nl
NGC6946~06 &$<$  3.12    & 98.13\p 1.31&$<$  1.19    &  8.06\p 0.60& 46.52\p 0.76&$<$  0.95    &  2.96\p 0.76& 54.94\p 0.81& 67.63\p 0.78\nl
NGC6946~07 &$<$  2.38    & 58.28\p 1.12&$<$  1.44    &  6.67\p 0.70& 28.11\p 0.83&$<$  0.78    &  2.01\p 0.48& 36.42\p 0.78& 44.61\p 0.72\nl
NGC6946~08 &$<$  2.32    & 57.43\p 1.06&$<$  1.18    &  6.75\p 0.54& 28.10\p 0.77&$<$  0.86    &  1.72\p 0.67& 41.55\p 0.93& 46.46\p 0.98\nl
NGC7793~00 &$<$  2.32    &  6.15\p 0.50&$<$  0.73    &  5.37\p 0.46&  4.92\p 0.66&$<$  0.54    &$<$  0.65    &  7.46\p 0.66&  7.67\p 0.85\nl
NGC7793~01 &  2.32\p 0.76& 11.43\p 1.05&$<$  1.60    &  4.32\p 0.51&  7.23\p 0.55&$<$  0.63    &  0.48\p 0.13& 10.96\p 0.60& 11.19\p 0.78\nl
NGC7793~02 &$<$  1.55    &  3.32\p 0.47&$<$  1.30    &  1.62\p 0.06&  2.23\p 0.60&$<$  0.54    &$<$  0.54    &  2.60\p 0.64&  4.50\p 0.89\nl
NGC7793~03 &$<$  2.79    & 11.28\p 0.82&$<$  1.17    &  4.92\p 0.63&  7.54\p 0.65&$<$  0.54    &  1.02\p 0.27&  9.06\p 0.58& 11.97\p 0.99\nl

\enddata
\tablecomments{\footnotesize Fluxes and their statistical uncertainties are listed in units of 10$^{-9}$~W~m$^{-2}$~sr$^{-1}$.  Calibration uncertainties are an additional $\sim$25\%.  The 3$\sigma$ upper limits are provided for nondetections in several key lines.  The nuclear targets are listed first followed by the extranuclear targets.}
\tablecomments{\footnotesize The nuclei for eight SINGS galaxies are not listed in this table.  No high resolution infrared spectral data were taken for the optical centers of Holmberg~I, Holmberg~II, M~81~Dwarf~A, M~81~Dwarf~B, IC~2574, DDO~154, NGC~3034 (M~82), and NGC~6822.}
\tablenotetext{a}{\footnotesize The infrared emission peaks outside of the field of view of the nuclear spectral maps.  The infrared center was thus observed as an extranuclear target, and the data from those observations are listed further down the table.}
\tablenotetext{b}{\footnotesize The Short-High maps are 45\arcsec$\times$33\arcsec\ instead of the standard $\sim$23\arcsec$\times$15\arcsec.}
\end{deluxetable}

\begin{figure}
 \plotone{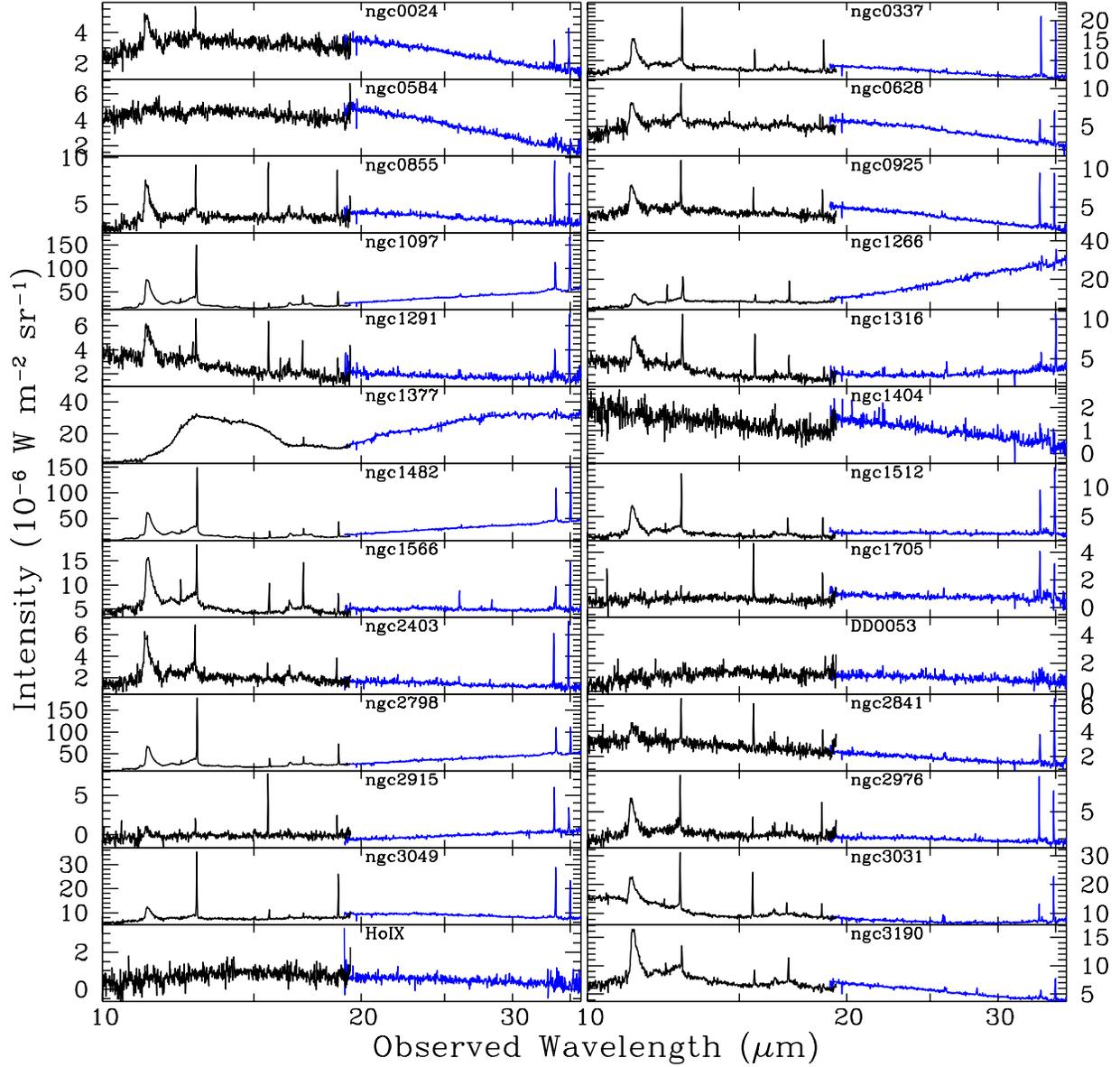}
 \caption{High resolution spectra for the SINGS nuclear and extranuclear targets.  The Short-High data span 10--19\m, while the Long-High data cover 19--37\m.  Note that the $y$-axis scalings differ for the lefthand and righthand columns of spectra.}
 \label{fig:spectra1}
\end{figure}

\setcounter{figure}{1}
\begin{figure*}
 \plotone{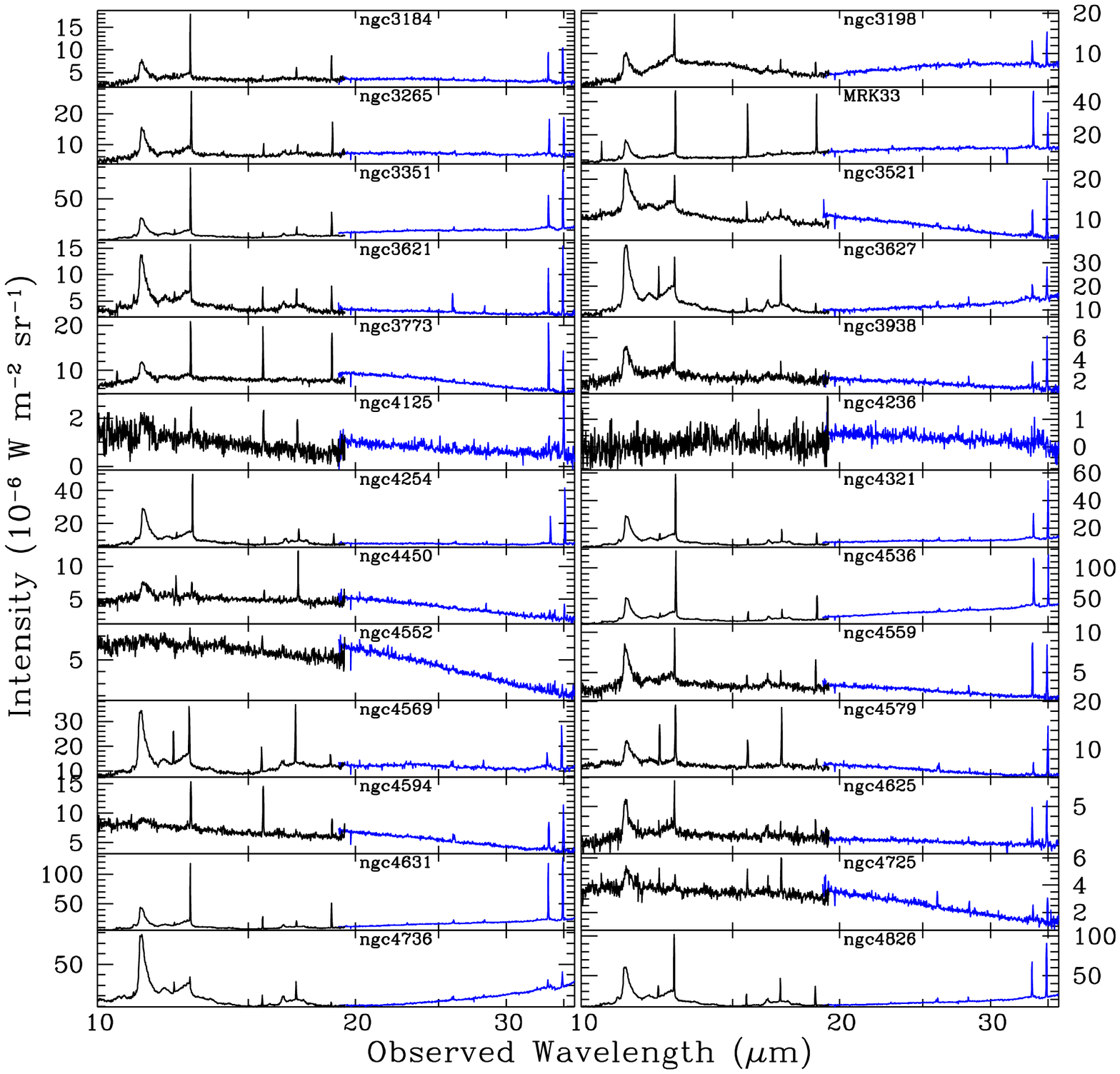}
 \caption{{\it Continued}}
 \label{fig:spectra2}
\end{figure*}

\setcounter{figure}{1}
\begin{figure}
 \plotone{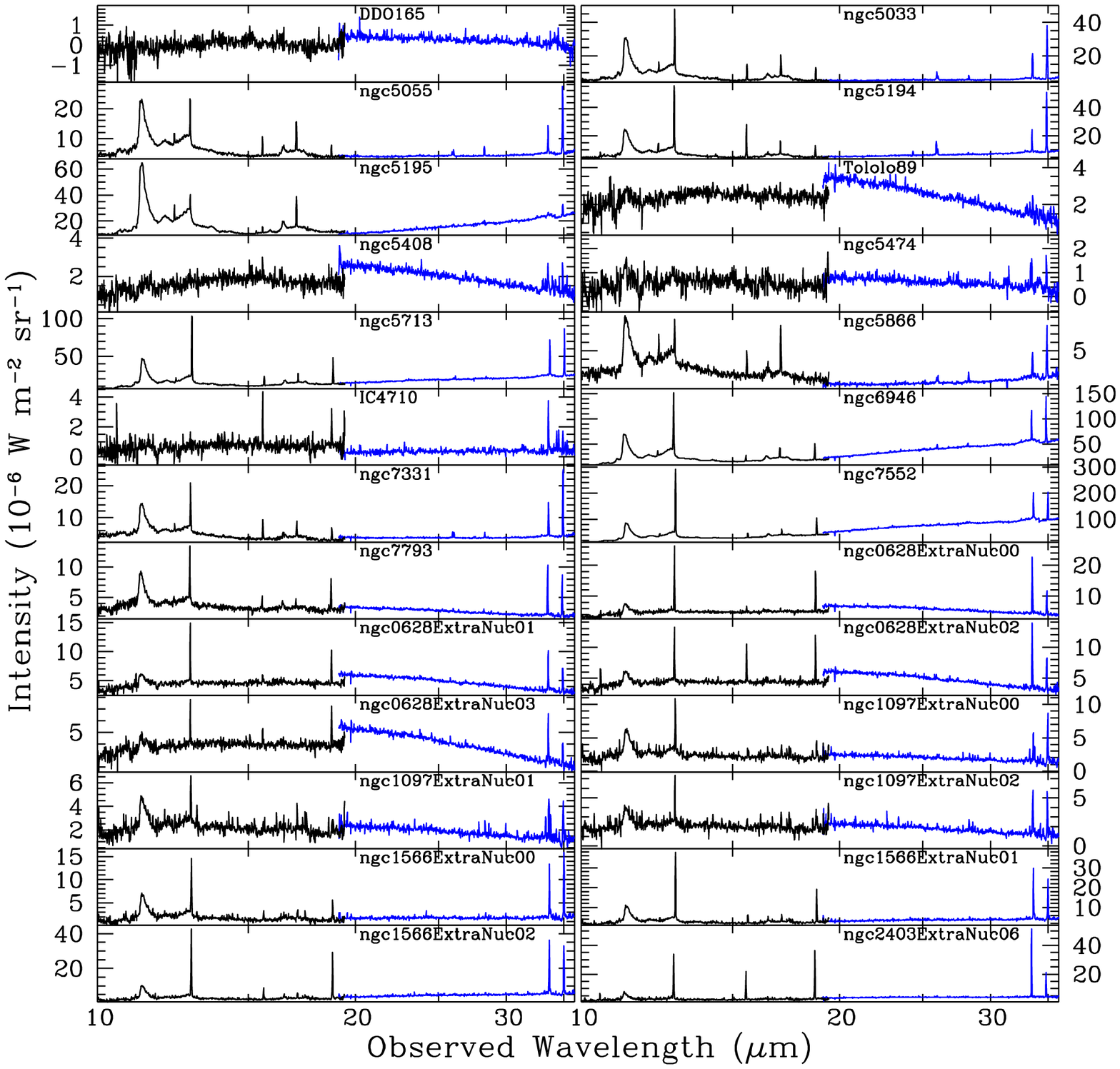}
 \caption{{\it Continued}}
 \label{fig:spectra3}
\end{figure}

\setcounter{figure}{1}
\begin{figure}
 \plotone{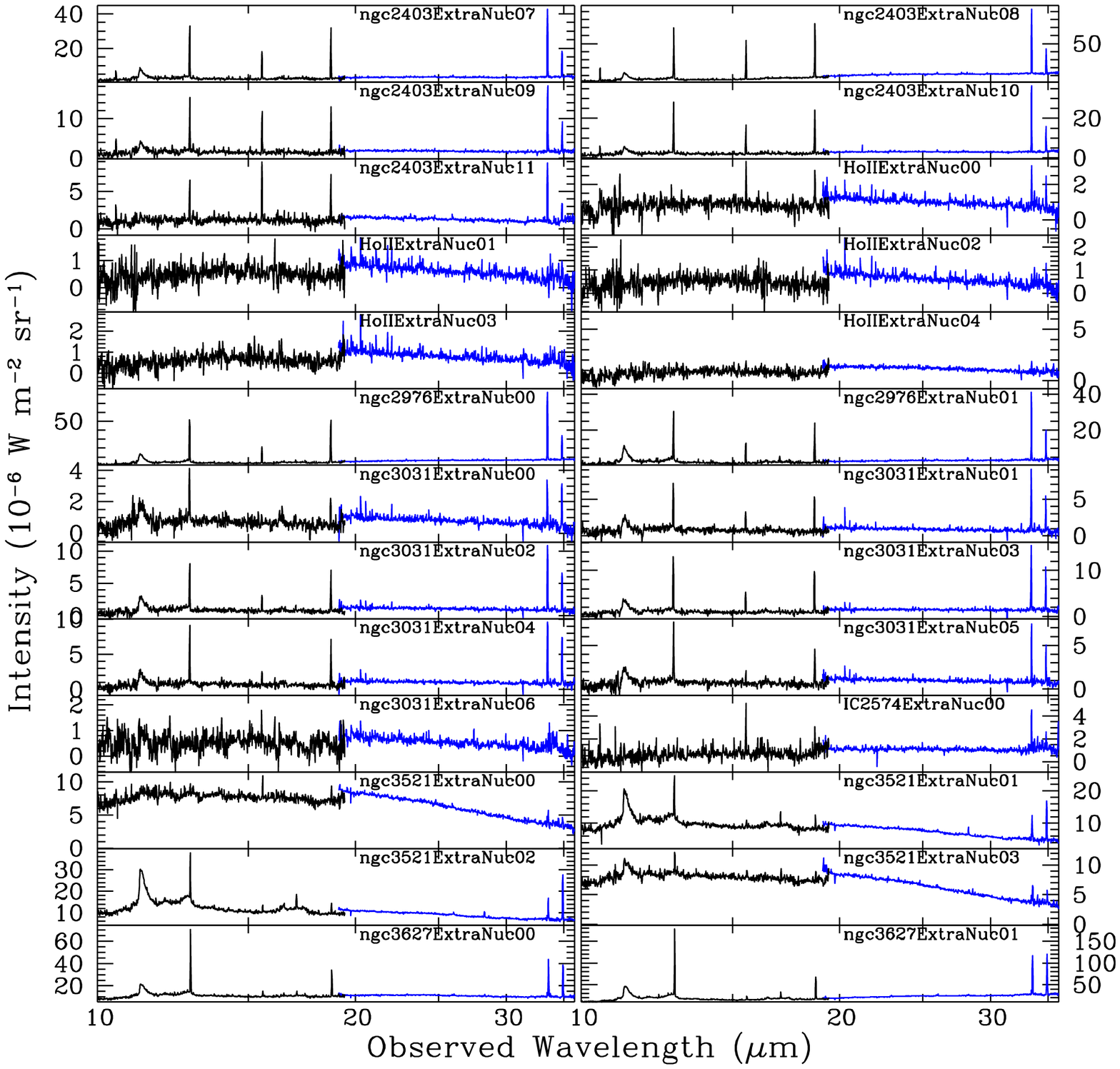}
 \caption{{\it Continued}}
 \label{fig:spectra4}
\end{figure}

\setcounter{figure}{1}
\begin{figure}
 \plotone{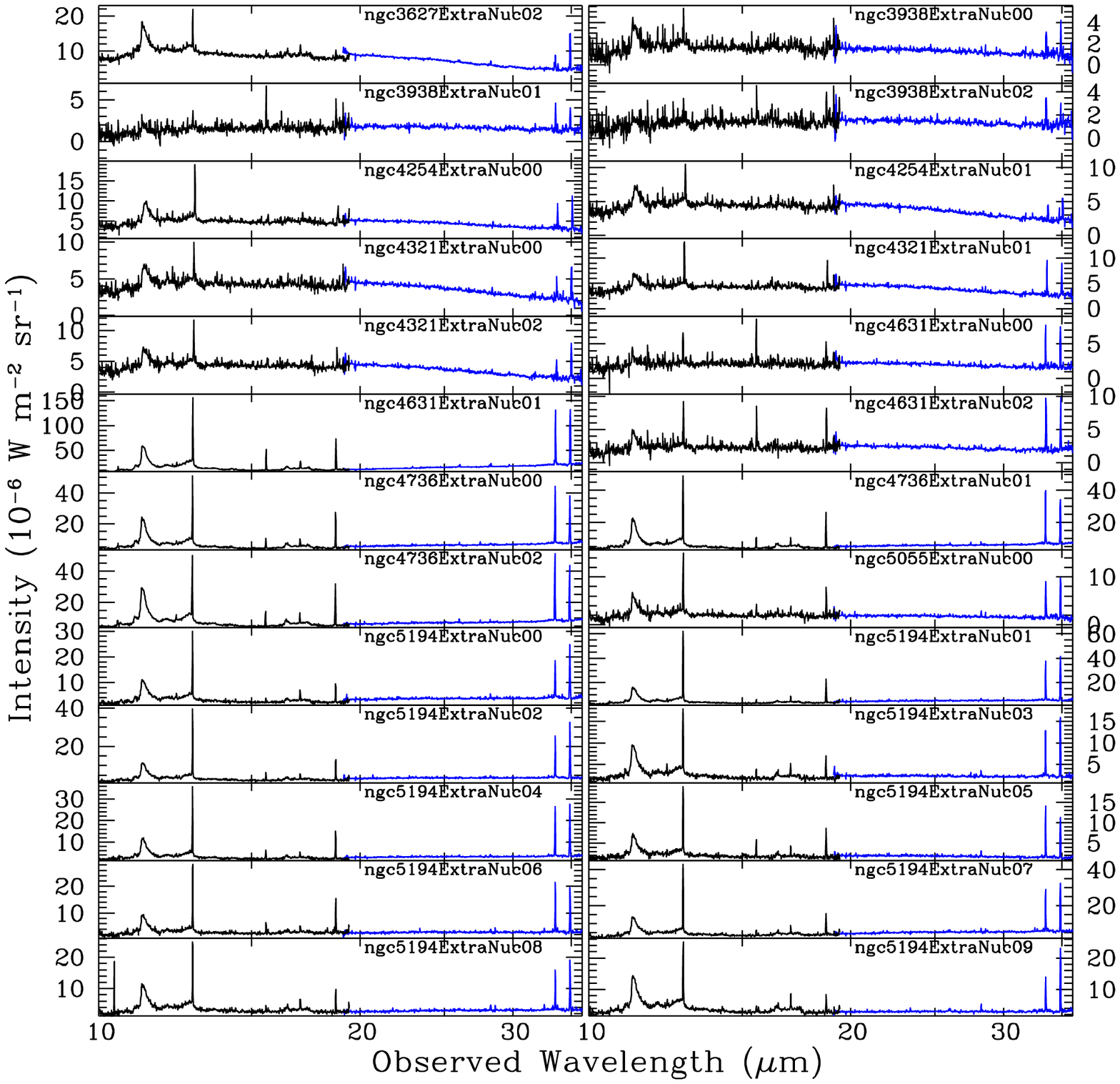}
 \caption{{\it Continued}}
 \label{fig:spectra5}
\end{figure}

\setcounter{figure}{1}
\begin{figure}
 \plotone{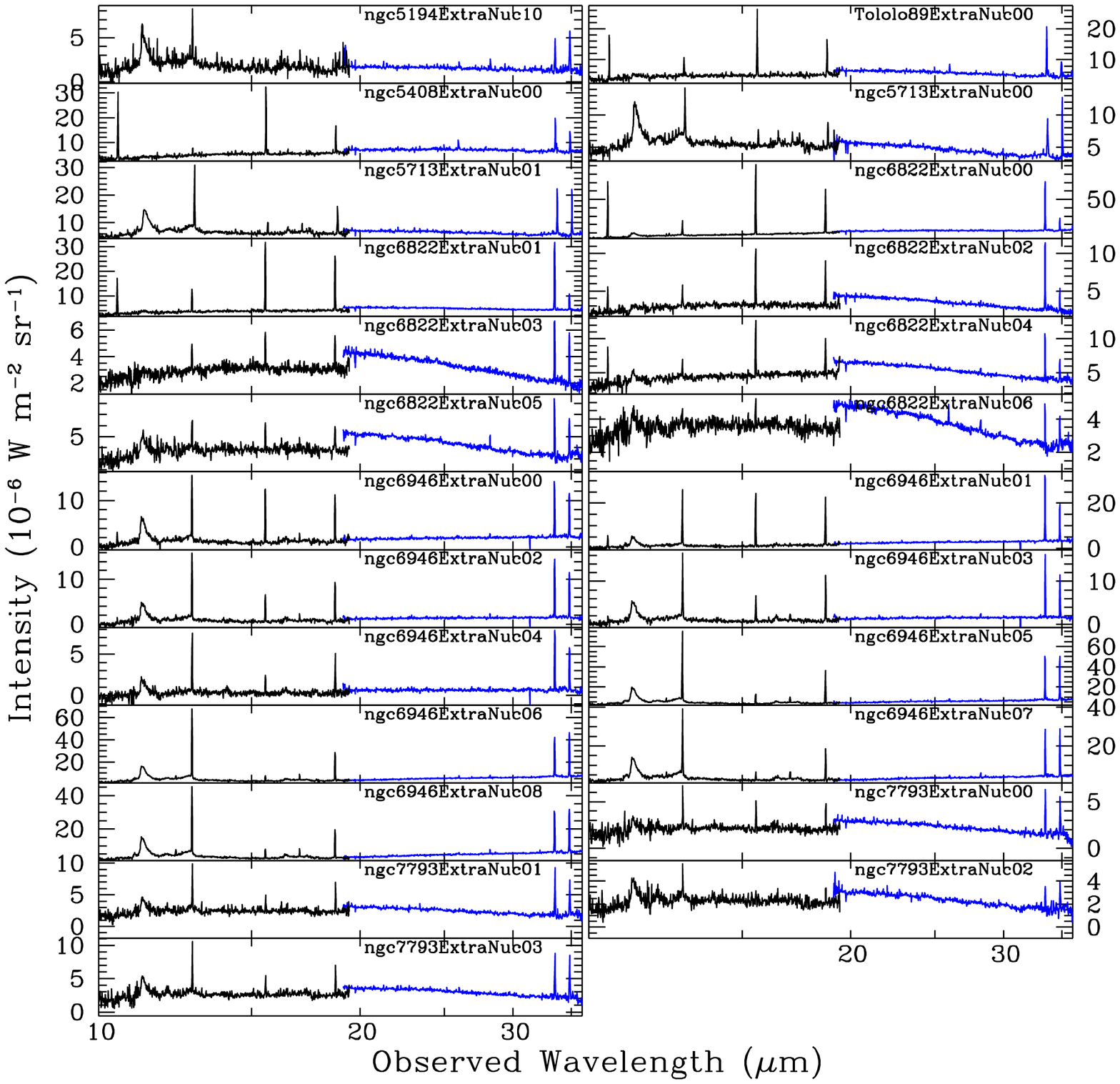}
 \caption{{\it Continued}}
 \label{fig:spectra6}
\end{figure}

\begin{figure}
 \plotone{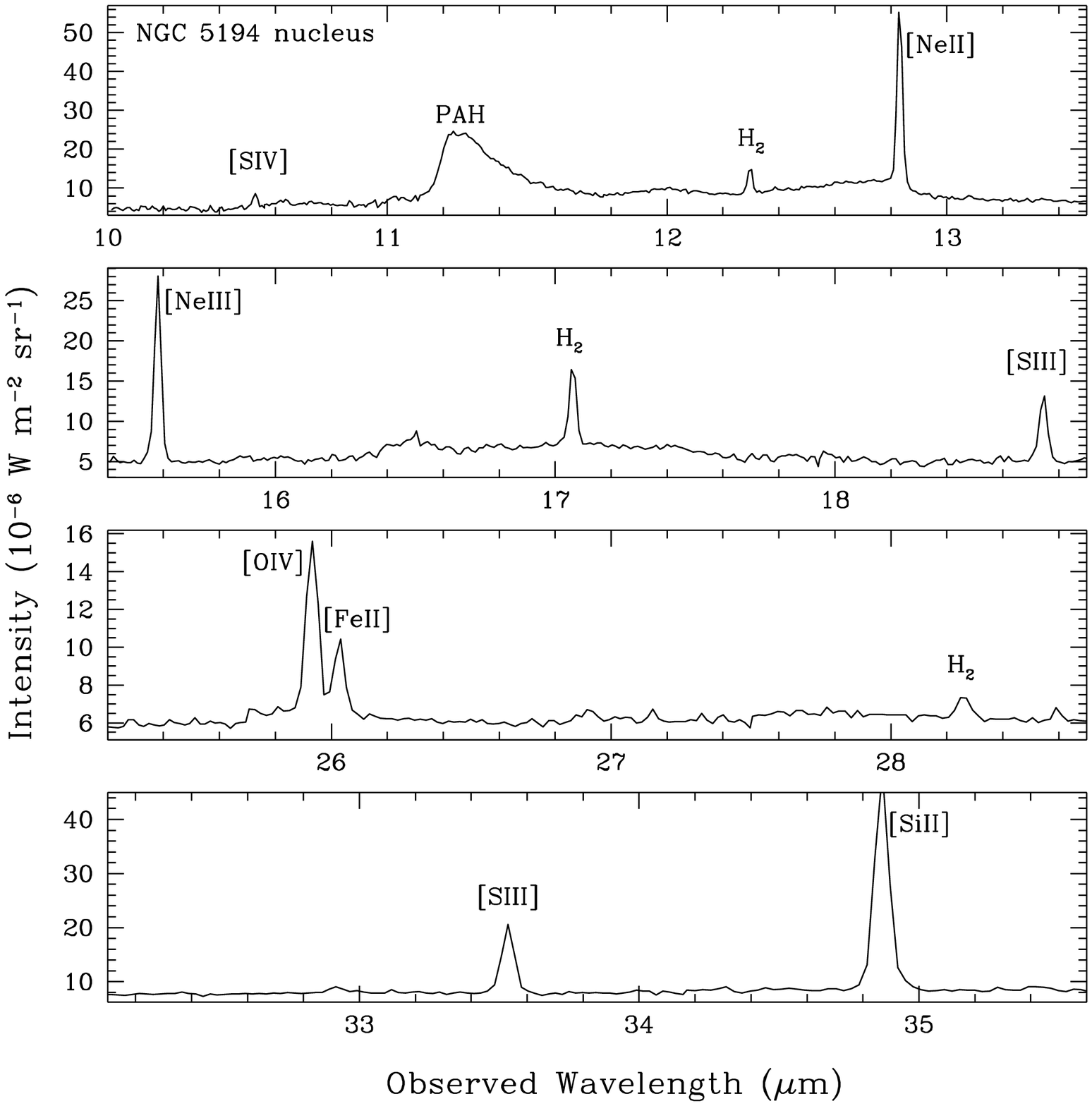}
 \caption{Close-up views of several emission features for the nucleus of NGC~5194.}
 \label{fig:blowup}
\end{figure}

\begin{figure}
 \plotone{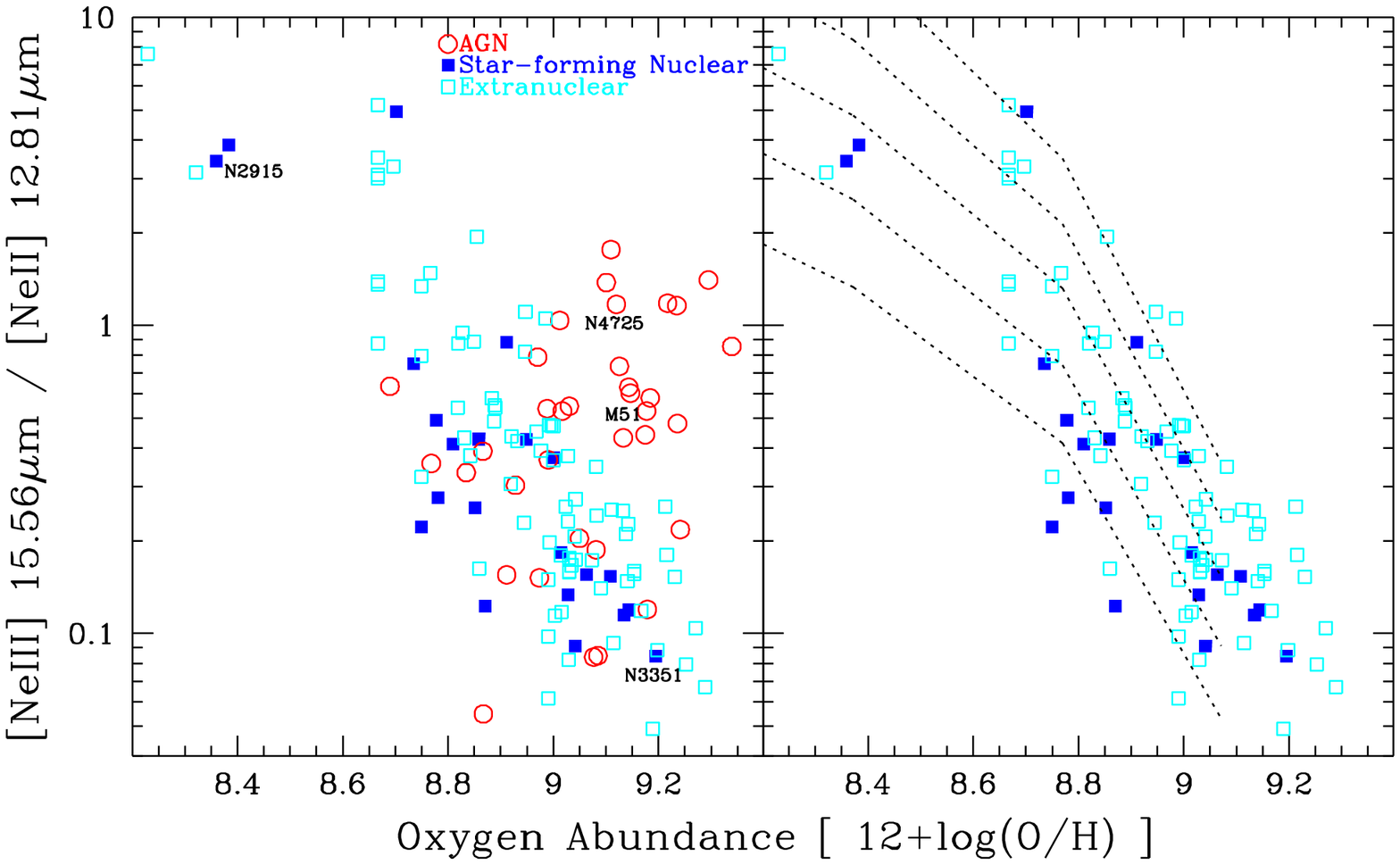}
 \caption{The ratio \NeIII 15.56\m/\NeII 12.81\m\ is displayed as a function of the oxygen abundance, as derived by Moustakas et al. (2009).  The abundances utilize the strong-line calibration of Kobulnicky \& Kewley (2004).  The righthand panel excludes the data from AGN.  Each extranuclear target is symbolized as an unfilled (cyan) square, regardless of its parent nucleus classification.  The series of dotted lines are predictions from photoionization models (Kewley et al. 2001), and in ascending order correspond to ionization parameters of $q=Uc=2\cdot10^7$, $4\cdot10^7$, $8\cdot10^7$, $1.5\cdot10^8$, and $3\cdot10^8$~cm~s$^{-1}$ (see \S~\ref{sec:hardness} for details).}
 \label{fig:Z}
\end{figure}

\begin{figure}
 \plotone{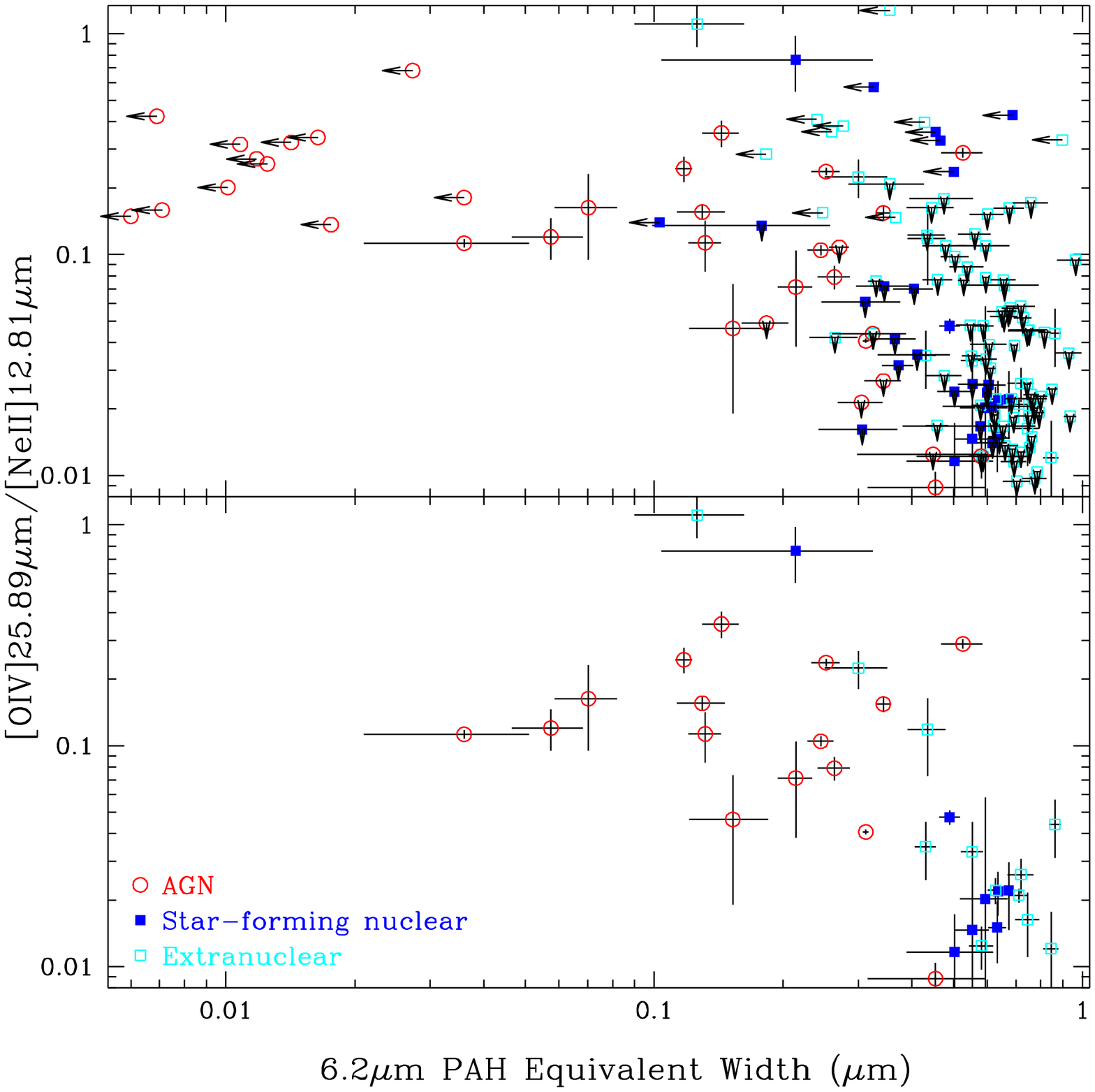}
 \caption{The ratio of \OIV~25.89\m/\NeII~12.81\m\ as a function of the 6.2\m\ PAH equivalent width.  The lower panel shows only sources with detections for \OIV~25.89\m, \NeII~12.81\m, and the 6.2\m\ PAH feature.}
 \label{fig:genzel}
\end{figure}

\begin{figure}
 \plotone{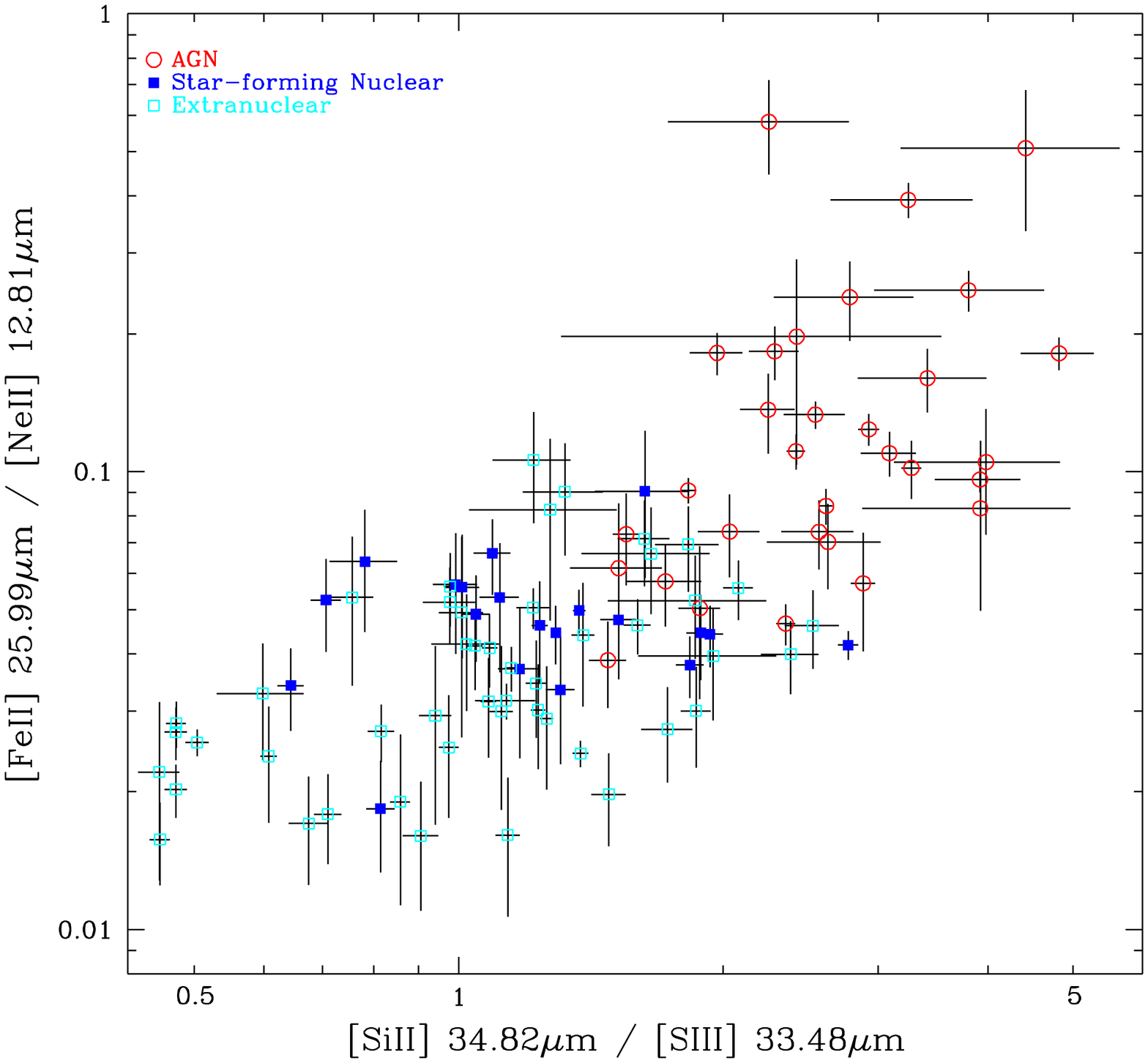}
 \caption{The ratio \FeII 25.99\m/\NeII 12.81\m\ is displayed as a function of the ratio \SiII 34.82\m/\SIII 33.48\m.}
 \label{fig:line_ratios}
\end{figure}

\begin{figure}
 \plotone{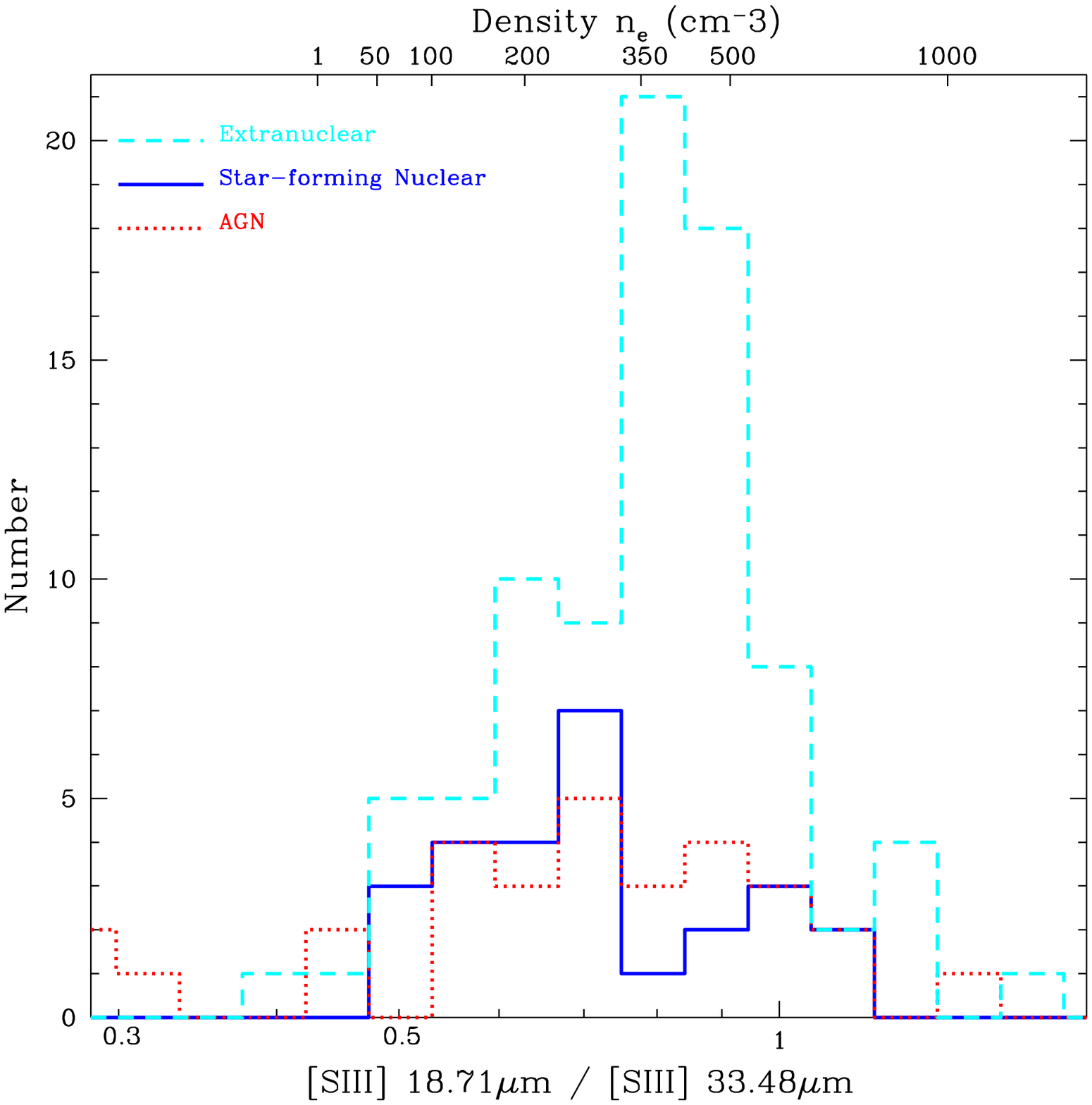}
 \caption{The distribution of nebular electron density as traced by the ratio \SIII 18.71\m/\SIII 33.48\m\ using $\sim$23\arcsec$\times$15\arcsec\ apertures.}
 \label{fig:density}
\end{figure}
\end{document}